\documentclass[usegraphicx,useAMS,usenatbib]{mn2e}
\usepackage{textcomp}
\usepackage[dvips]{graphicx}

\usepackage{pdflscape}
\usepackage{afterpage}
\usepackage{capt-of}

\usepackage{lipsum}

\newcommand{\teff}{T_{\rm eff}}

\newcommand{\bz}{$\langle B_z \rangle$}

\newcommand{\vsini}{$v \sin i$}
\newcommand{\kms}{km\,s$^{-1}$}

\newcommand{\mdot}{$\dot{M}$}
\newcommand{\vinf}{$v_\infty$}

\newcommand{\msun}{$M_\odot$}

\newcommand{\ra}{$R_{\rm A}$}
\newcommand{\rk}{$R_{\rm K}$}

\title[Magnetic field of HD~184927]
  {The surface magnetic field and chemical abundance distributions of the B2V helium-strong star HD~184927}
\author[I.~Yakunin et al.]
  {I.~Yakunin$^1$\thanks{Mail to:elias@sao.ru},
  G.~Wade$^2$, D.~Bohlender$^3$, O.~Kochukhov$^4$, W.~Marcolino$^5$,
  M.~Shultz$^{2,3,10}$, 
  \newauthor 
  D.~Monin$^6$,
  J.~Grunhut$^{7}$,
  T.~Sitnova$^8$,
  V.~Tsymbal$^9$
    and the MiMeS Collaboration \\
  $^1$Special Astrophysical Observatory, Nizhniy Arkhyz, Russia 369167\\
  $^2$Department of Physics, Royal Military College of Canada, Kingston, Ontario, Canada K7K 7B4\\
  $^3$National Research Council of Canada, Herzberg
Astronomy and Astrophysics Program,\\ 5071 West Saanich Road, Victoria, BC, V9E 2E7, Canada\\
  $^4$Department of Physics and Astronomy, Uppsala University, SE-751 20, Uppsala, Sweden\\
  $^5$Universidade Federal do Rio de Janeiro, Observatorio do Valongo. Ladeira Pedro Antônio, 43, CEP 20080-090, Rio de Janeiro, Brazil\\
  $^6$Dominion Astrophysical Observatory, NRC-HIA, 5071 West Saanich Road, Victoria, BC, V9E 2E7, Canada\\
  $^7$European Southern Observatory, Karl-Schwarzschild-Str. 2, D-85748 Garching, Germany\\
  $^8$Moscow State University, Moscow, 119991 Russia\\
  $^9$Tavrian National University, Vernadskiys Avenue 4, Simferopol, Crimea, 95007\\
  $^{10}$European Organisation for Astronomical Research in the Southern Hemisphere, Casilla 19001, Santiago 19, Chile\\  
  }
\date{MNRAS Accepted}

\pagerange{\pageref{firstpage}--\pageref{lastpage}} \pubyear{2014}

\begin{document}

\label{firstpage}

\maketitle

\begin{abstract}
A new time series of high-resolution Stokes $I$ and $V$ spectra of the magnetic B2V star HD 184927 has been obtained in the context of the Magnetism in Massive Stars (MiMeS) Large Program with the ESPaDOnS spectropolarimeter at the Canada-France-Hawaii Telescope and dimaPol liquid crystal spectropolarimeter at 1.8-m telescope of Dominion Astrophysical Observatory. We model the optical and UV spectrum obtained from the IUE archive to infer the stellar physical parameters. Using magnetic field measurements we derive an improved rotational period of $9.53102\pm0.0007$~d. We infer the longitudinal magnetic field from lines of H, He and various metals, revealing large differences between the apparent field strength variations determined from different elements. Magnetic Doppler Imaging using He and O lines yields strongly nonuniform surface distributions of these elements. We demonstrate that the diversity of longitudinal field variations can be understood as due to the combination of element-specific surface abundance distributions in combination with a surface magnetic field that is comprised of dipolar and quadrupolar components. We have reanalyzed IUE high resolution spectra, confirming strong modulation of wind-sensitive C~{\sc iv} and S~{\sc iv} resonance lines. However, we are unable to detect any modulation of the H$\alpha$ profile attributable to a stellar magnetosphere. We conclude that HD 184927 hosts a centrifugal magnetosphere ($\eta_*\sim 2.4^{+22}_{-1.1}\times 10^{4}$), albeit one that is undetectable at optical wavelengths. The magnetic braking timescale of HD 184927 is computed to be $\tau_J = 0.96$ or $5.8$ Myr. These values are consistent with the slow rotation and estimated age of the star.
\end{abstract}


\section{Introduction}

Magnetic early B-type stars were first identified in the 1970s with the detection of a strong, organized magnetic field in the peculiar B2V star $\sigma$~Ori E \citep{lb1978}. Following this discovery, systematic surveys (e.g. \citealt{bl1979, bohl1987}) demonstrated that the helium-strong B stars are a class of magnetic stars, with surface magnetic field strengths on the order of kG.  More recently, surveys with modern high-resolution spectropolarimeters have detected magnetic fields in other early B stars, some of which are helium-peculiar stars (e.g. \citealt{neiner03, oks10, grun2012}), and others which lack such signature chemical peculiarities (e.g. \citealt{henrichs13, petit11}).

Magnetism in early B stars modifies their spectroscopic and physical properties in important ways. In the cooler early B stars with $\teff \sim 22-25$~kK, microscopic chemical diffusion produces strong vertical and horizontal chemical abundance gradients. The local accumulation of chemical elements can introduce important changes in the atmospheric structure of the star, modifying the optical and UV continuum and profiles of Balmer lines, introducing apparent departures from LTE in the line spectra of various elements, and producing line profile variability. The wind-magnetic field interaction alters the wind flux and geometry, introducing emission in Balmer lines, and changing the bulk and internal rotation of the star (e.g. \citealt{petit2013}). The diverse spectral and rotational properties of the magnetic B stars indicate that they occupy a sensitive region of parameter space that has much to tell us concerning diffusion, mixing, magnetic wind confinement and rotational braking.

HD 184927 ($=$V\, 1671 Cyg$=$HIP 96362) is a B2V helium-strong star. Strong lines of He were first identified by \cite{bond1970}, who noted a similarity to $\sigma$ Ori E, the prototypical He-strong star. \cite{walborn75} and \cite{bond76} reported significant variability in the strengths of the He lines, as well as photometric variability in the Str\" omgren $u$ band. The He equivalent widths and the photometry were found to vary with a period of 9.48~d, interpreted as the rotational period of the star according to \citeauthor{stibbs50}'s (\citeyear{stibbs50}) oblique rotator model. (This period was later refined to  $9. 536\pm 0.05$~d  by \cite{lm70} using additional He equivalent width measurements.)

\cite{wade97} analyzed the rotation and magnetic field geometry of HD 184927. Using measurements of the longitudinal magnetic field, He line strength and $u$ band magnitude, they improved the rotational period determination to $9.52961\pm 0.00731$~d. When phased with this period, their H$\beta$ magnetic field measurements described an approximately sinusoidal variation from about -0.7 to +1.8~kG. They fit spectral absorption lines of C~{\sc ii} $\lambda 6582.9$ and derived $v\sin i=14.5\pm 2.5$~km/s. Using $\teff=22.5\pm 0.6$\,K and $\log g=3.80\pm 0.05$ as derived by \cite{higlee74}, they inferred a stellar mass of $10\pm 1~M_\odot$ and a radius of $R=6.6\pm 0.8~R_\odot$. Using the measured $v\sin i$, rotational period and longitudinal field variation, they inferred a surface magnetic field geometry with $i=25\pm 5\degr$, $\beta=78\pm 3\degr$ and $B_{\rm d}=9.7-13.7$~kG.

HD 184927 is a significant object for several reasons. With a rotational period of $\sim$10~d, it is one of the most slowly-rotating He strong stars. As a consequence of its slow rotation, its spectral lines are relatively sharp. This allows for a detailed analysis of its line spectrum for the purpose of atmospheric parameter and abundance determination. In addition, the combination of sharp lines and strong magnetic field produce clear Stokes signatures in individual spectral lines that are suitable for Magnetic Doppler Imaging (e.g. \citealt{kp2002}). Finally, HD 184927 exhibits significant emission and variability of its ultraviolet (UV) C~{\sc iv} and Si~{\sc iv} resonance lines \citep{barker1982}, which directly probes the interaction of the stellar wind and magnetic field. Puzzlingly, however, it shows no sign of the H$\alpha$ emission that often accompanies this interaction. For these reasons, HD 184927 was included in the Targeted Component sample of the Magnetism in Massive Stars (MiMeS) project \citep{mimes}.

In this paper we report the data acquired and use it together with the data retrieved from archives to analyze the object. In Sect. 2 we describe observational data used in our study. In Sect. 3 we describe the measurement of the longitudinal magnetic field from the ESPaDOnS spectra, from individual lines and LSD profiles. In Sect. 4 we discuss the refinement of the period of rotation and introduce a new rotational ephemeris for the star. In Sect. 5 we describe in detail the variations of spectral lines and the magnetic field inferred from various chemical elements. Sect. 6 reports modelling of the spectrum of the star using NLTE TLUSTY models. Sect. 7 is dedicated to Magnetic Doppler Imaging of the star based on the He~{\sc i} $\lambda$6678 line and the IR oxygen triplet ($\lambda$7772+7774+7775). In Sect. 8 we derive constraints on the magnetospheric properties of HD~184927. We summarise our results and conclude in Sect. 9.


\section{Observations}

The spectra used in this study were obtained from two main sources - the ESPaDOnS spectropolarimeter located at the Canada-France-Hawaii Telescope \citep[e.g.][]{silvester12} and the dimaPol liquid crystal spectropolarimeter installed in the DAO 1.8-m telescope \citep{monin12}.

\subsection{ESPaDOnS spectra}
\label{sect:espadons}
Twenty-eight Stokes $V$ spectra of HD 184927 were obtained with the ESPaDOnS spectopolarimeter at the Canada-France-Hawaii Telescope (CFHT) between 2008 August 20 and 2012 June 27. Each spectropolarimetric sequence consisted of four individual subexposures, each of 500 or 600 s duration, taken in different configurations of the polarimeter retarders. From each set of four subexposures we derived Stokes $I$ and Stokes $V$ spectra in the wavelength range 3670 -- 10000 \AA\ following the double-ratio procedure described by e.g. \citet{bag2009}, ensuring that all spurious signatures are removed to first order. Diagnostic null polarization spectra (labeled $N$) were calculated by combining the four subexposures in such a way that polarization cancels out, allowing us to verify that no spurious signals are present in the data (for more details on the definition of $N$, see e.g. Donati et al. 1997). All spectra were reduced at CFHT using the Upena pipeline feeding the Libre-ESpRIT code \citep{d1997}. Three poorly-exposed spectra acquired on the same night (HJDs 2454756.7331, 2454756.7841 and 2454756.8115) were combined to increase signal-to-noise ratio. The resulting spectra are characterized by a resolving power $R\sim65\, 000$ and peak S/N ratios from 113-613 per 1.8~km/s pixel at about 550 nm (see Table \ref{esp:data:v}).

\begin{figure}
\centering
 \includegraphics[scale=0.47]{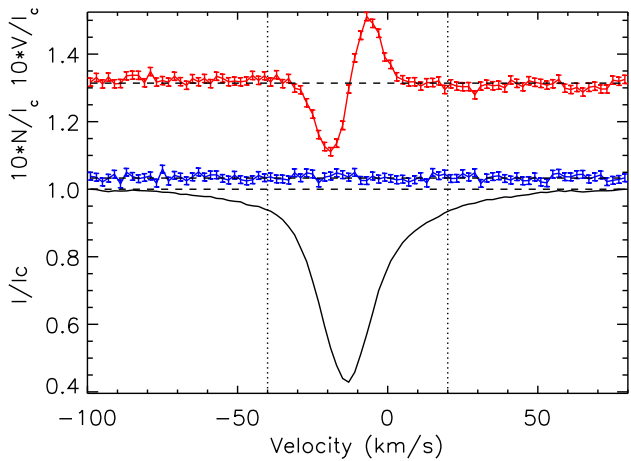}
 \caption{Mean LSD Stokes $V$\,(top), diagnostic null\,(middle) and Stokes $I$\,(bottom) of HD~184927 from 22 Aug 2008. The $V$ and $N$ profiles are expanded by the indicated factor and shifted upwards. The error bars represent the $1\sigma$ uncertainties. The integration limits used to measure the longitudinal field in the ESPaDOnS LSD profiles are indicated by the dotted lines.}
 \label{fig:profile}
\end{figure}

We then applied the Least-Squares Deconvolution (LSD) procedure \citep{d1997} using the implementation of \citet{koch2010} to obtain mean Stokes $I$, $V$ and $N$ profiles, improving the S/N ratio of our polarimetric measurements by approximately a factor of 3. At the beginning our LSD mask contained lines from ``extract stellar'' output from the Vienna Atomic Line Database \citep[VALD][]{2000BaltA...9..590K} with $\teff=22000$\,K and $\log g=4.0$ corresponding to our star and intrinsic line depths greater then 5\% of the continuum. Then we removed all the lines that were blended with hydrogen and that were too weak to be visible in the spectrum. After removing the broadest helium lines and their blends, the resulting ``full mask'' contains 282 lines, most of which are metal lines and weak helium lines. The resulting S/N ratios of the LSD profiles, normalized according to weights computed from $\lambda=500$~nm, $g=1.2$ and $d=0.2$, range from 500 to 7500. Examples of final LSD profiles are shown in Fig. \ref{fig:profile}.

We also obtained 14 Stokes Q/U spectra with ESPaDOnS with the aim of analyzing the transverse field of HD~184927. We did not perform LSD procedure to these data, instead focusing on analysis of the linear polarisation in the individual spectral lines. The log of Stokes Q/U observations is reported in Table \ref{esp:data:qu}. The analysis of these data is briefly described at the end of the Section \ref{MDI}.

\begin{table}
\caption{ESPaDOnS Stokes V data used in the analysis. Columns are Heliocentric Julian Date (HJD), phase according to the new ephemeris (Eq. (\ref{ephem})), the exposure time, peak SNR per 1.8 \kms\,pixel for individual spectra, SNR per 1.8 \kms\,pixel for the LSD profiles obtained using the full mask, and integration range used to compute magnetic field.}
\begin{tabular}{c|c|c|c|c|c|c|c}
\hline
\hline
HJD    & Phase  & Exp. & Peak  & SNR & Int. range\\
	   &            & (s)      &    SNR       &(LSD)& \kms \\
\hline
2454667.9326& 0.031 & 2000 & 472 &5689& -47/31\\
2454667.9593& 0.034 & 2000 & 507 &5949& -47/31\\
2454698.9072& 0.281 & 2000 & 565 &6645& -39/23\\
2454699.8416& 0.379 & 2000 & 567 &6740& -37/21\\
2454699.9123& 0.386 & 2000 & 563 &6684& -37/21\\
2454700.8352& 0.483 & 2000 & 421 &4931& -37/19\\
2454700.9051& 0.491 & 2000 & 303 &3599& -37/19\\
2454754.7586& 0.141 & 2000 & 351 &4232& -45/29\\
2454754.7858& 0.144 & 2000 & 487 &5784& -45/29\\
2454756.7331& 0.348 & 2000 & 95  &1057& -37/19\\
2454756.7841& 0.353 & 2000 & 61  &484 & -37/19\\
2454756.8115& 0.356 & 2000 & 164 &1771& -39/21\\
2454759.7313& 0.663 & 2000 & 553 &6709& -39/21\\
2454954.0990& 0.056 & 2400 & 577 &6980& -47/31\\
2454954.9434& 0.144 & 2400 & 601 &7270& -47/29\\
2454955.9478& 0.250 & 2400 & 471 &5619& -41/23\\
2454959.0524& 0.576 & 2400 & 356 &4326& -37/19\\
2454963.1008& 0.000 & 2400 & 508 &6070& -47/31\\
2455019.1256& 0.879 & 2400 & 529 &6237& -45/29\\
2455251.1717& 0.225 & 2000 & 528 &6350& -41/25\\
2455253.1548& 0.433 & 2000 & 466 &5638& -37/19\\
2455258.1328& 0.955 & 2000 & 323 &3821& -47/29\\
2455394.7707& 0.291 & 2000 & 508 &6216& -39/23\\
2455522.6867& 0.712 & 2000 & 613 &7142& -41/23\\
2455879.7358& 0.174 & 2000 & 602 &7461& -45/29\\
2456102.9978& 0.599 & 2000 & 468 &5772& -37/19\\
2456104.0072& 0.705 & 2000 & 439 &5563& -39/21\\
2456105.9222& 0.906 & 2000 & 518 &6484& -45/29\\
\hline
\end{tabular}   
\label{esp:data:v}
\end{table}

\begin{table}
\caption{ESPaDOnS Stokes Q/U data used in the analysis. Columns are Heliocentric Julian Date (HJD), phase according to the new ephemeris (Eq. (\ref{ephem})), the exposure time, peak SNR per 1.8 \kms\,pixel for individual spectra, and Stokes parameter.}
\begin{tabular}{c|c|c|c|c}
\hline
\hline
HJD    & Phase  & Exp. & Peak  & Stokes \\
       &        & (s)  & SNR   & \\
\hline
2454667.9845& 0.037 & 2000 & 544 & Q \\
2454668.0098& 0.039 & 2000 & 542 & U \\
2454668.0373& 0.042 & 2000 & 272 & Q \\
2454668.0624& 0.044 & 2000 & 354 & U \\
2454698.9333& 0.284 & 2000 & 659 & Q \\
2454698.9591& 0.286 & 2000 & 635 & U \\
2454699.7896& 0.374 & 2000 & 705 & Q \\
2454699.8155& 0.376 & 2000 & 724 & Q \\
2454699.9382& 0.389 & 2000 & 665 & U \\
2454699.9641& 0.392 & 2000 & 655 & U \\
2454700.7840& 0.478 & 2000 & 665 & Q \\
2454700.8096& 0.481 & 2000 & 655 & Q \\
2454700.9305& 0.493 & 2000 & 665 & U \\
2454700.9558& 0.496 & 2000 & 655 & U \\
\hline
\end{tabular}   
\label{esp:data:qu}
\end{table}

\subsection{DAO observations}

The 1.8-m Dominion Astrophysical Observatory (DAO) Plaskett telescope and $dimaPol$ \citep{monin12} spectropolarimeter were used to obtain 11 observations of HD\,184927. Left- and right-circularly polarized spectra with a resolution of about 10,000 covering an approximately 250\,\AA\,wide spectral window centred on the H$\beta$ line were recorded with the SITe-2 CCD. A typical two-hour observation sequence consisted of 12 sub-exposures of 600\,s each, with 60 switches of the liquid crystal quarter-wave plate during each sub-exposure. See Table \ref{dima:data} for details. More detailed information about the instrument and the observing and data reduction procedures is provided by \cite{monin12}.

\begin{table}
\caption{Data obtained from dimaPol. Columns are Heliocentric Julian Date, phase according to the new ephemeris (Eq. \ref{ephem}), SNR for individual spectra, the exposure time, derived longitudinal field and error from $H\beta$ and He~{\sc ii}\ $\lambda$4922 lines. }
\begin{tabular}{p{1.5cm}|p{0.5cm}|p{0.5cm}|p{0.5cm}rrr}
\hline
\hline
HJD  & Phase    & SNR & Exp. &H$\beta$ field&He~{\sc i} field\\
 & & &(min) &  Be (G)  & Be (G) \\
\hline
2455699.928 & 0.255 & 289& 110& $   838\pm210 $ & $   82  \pm214$\\
2455700.923 & 0.314 & 309& 120& $-257  \pm137 $ & $ -236  \pm239$\\
2455701.935 & 0.366 & 388& 110& $ 159  \pm228 $ & $ -179  \pm205$\\
2455705.902 & 0.614 & 386& 120& $1755  \pm192 $ & $ 1520  \pm120$\\
2455716.805 & 0.276 & 433& 120& $1827  \pm184 $ & $ 1481  \pm206$\\
2455717.832 & 0.261 & 348& 180& $1630  \pm214 $ & $  652  \pm178$\\
2455721.890 & 0.420 & 290& 120& $ 483  \pm304 $ & $ -329  \pm273$\\
2455733.809 & 0.987 & 407& 180& $1961  \pm238 $ & $  976  \pm171$\\
2455761.907 & 0.324 & 387& 120& $1402  \pm130 $ & $  936  \pm170$\\
2455782.863 & 0.183 & 313& 120& $2141  \pm268 $ & $ 1294  \pm231$\\
2455789.793 & 0.310 & 458& 160& $ 786  \pm178 $ & $  364  \pm104$\\
\hline
\end{tabular}
\label{dima:data}
\end{table}

\subsection{Ultraviolet spectra}

We have also downloaded 32 International Ultraviolet Explorer (IUE) observations from the IUE archive. The spectra were obtained with the short wavelength prime (SWP) camera in high-dispersion mode (MXHI). The data were processed with the New Spectral Image Processing System (NEWSIPS), providing wavelengths, absolute calibrated fluxes, data quality flags, and noise estimates. There are no significant differences between the NEWSIPS and INES data in the case of HD~184927. Spectra were normalized to pseudo-continuum regions close to the resonance lines of interest; as the so-called `iron forest' depresses the true continuum by a substantial amount, this procedure does not derive the true continuum. However, as we are concerned here primarily with the variability of the lines, rather than detailed quantitative modeling, this normalization procedure is adequate because continuum UV flux variability is usually no more than a few percent\,(e.g. \cite{smithgroote2001}, Fig. 4).
\section{Magnetic field measurements} \label{sect:espmag}

For $dimaPol$ spectra longitudinal magnetic field values, \bz, were obtained by measuring the Zeeman shift between the two opposite circular polarizations in the core of the H$\beta$ and He~{\sc i} $\lambda$4922 lines using a Fourier cross-correlation technique. The observed shift is proportional to the longitudinal field with a scaling factor of 6.8\,kG per pixel for H$\beta$ and 6.6\,kG per pixel 
for the helium line.

The measured longitudinal field ranges from -257 to +2141 G (H$\beta$), and from -329 to +1520 G (He~{\sc i}), with typical uncertainties of about 180 G.  The magnetic field measurements determined from each line can be found in Table \ref{dima:data}.

Longitudinal magnetic field measurements were obtained from the ESPaDOnS spectra in several ways.

First, the longitudinal magnetic field was inferred from each LSD Stokes $I/V$ profile set by computing the first-order moment of  Stokes $V$ normalised to the equivalent width of Stokes $I$:

\begin{equation}
\langle B_z\rangle=-2.14\times 10^{11}\ \frac{{\displaystyle \int (v-v_{\rm 0}) V(v)\ dv}}{\displaystyle {\lambda g c\ \int [1-I(v)]\ dv}}
\label{bzeqn}
\end{equation}

\noindent (see \citealt{wade2000} for details).  In Eq.~(\ref{bzeqn}) $V(v)$ and $I(v)$ are the $V/I_{\rm c}$ and $I/I_{\rm c}$ LSD profiles, respectively. $c$ is the speed of light. The wavelength $\lambda$ is expressed in nm and the longitudinal field \bz\ is in gauss. The wavelength and Land\'e factor $g$ used to compute the longitudinal field are the same values employed to normalize the LSD profiles (i.e. 500 nm and 1.2, respectively). $v_{\rm 0}$ corresponds to the centre of gravity of Stokes V. Uncertainties were calculated by propagating the formal uncertainties of each LSD spectral pixel through Eq. (\ref{bzeqn}). A detailed descrition of the procedure is provided by \cite{silvester2009}.

The integration range was automatically adapted to encompass the observed span of the Stokes $I$ profile, as well as the full profile range of the Stokes $V$ signature. Typically the range borders correspond roughly to the first and last points where the flux is equal to 85 per cent of continuum (see Fig. \ref{fig:profile} and Table \ref{esp:data:v}). Decreasing the integration range has an important impact on the derived longitudinal field, since part of the line is being ignored. For larger integration ranges, this adds continuum noise to the measurement, increasing the error bar but no significantly changing the derived measurements. The longitudinal field measured using this method ranges from -216 to +1027 G with uncertainties of 15-20~G.

\begin{figure}
\centering
 \includegraphics[scale=0.65]{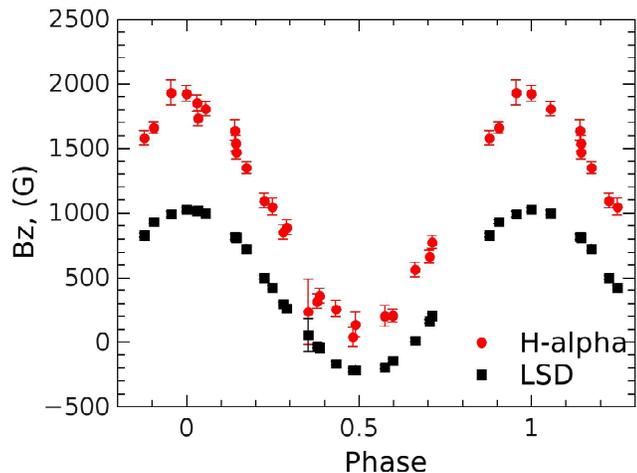}
 \caption{Longitudinal magnetic field computed using H$\alpha$ line (red) versus that obtained from LSD profiles(black).}
 \label{fig:halsd}
\end{figure}

We also used Eq.~(\ref{bzeqn}) to measure \bz\ from the cores of the H$\alpha$ and H$\beta$ Balmer lines. Again, the integration range was chosen to include the full Stokes $V$ profile. We integrated within a region of approximately $\pm 55$~km/s around line centre. The longitudinal field measurements obtained from the H lines range from about 0 to 2000 G, with typical uncertainties of 70~G (for H$\alpha$) and 120~G (for H$\beta$). The range of the longitudinal field variation measured from these lines is consistent with that obtained with dimaPol, and also consistent with that reported in the literature \citep{wade97}. However, it is significantly different from that measured from the LSD profiles. In particular, the maximum longitudinal field measured from the H lines is over twice as large as that measured from the LSD profiles (see Fig. \ref{fig:halsd}). In addition, an examination of the LSD Stokes $V$ signatures reveals that they reverse polarity, consistent with the inferred negative values of \bz. In contrast, the ESPaDOnS H line Stokes $V$ signatures maintain the same sign. Such discrepancies between the longitudinal magnetic field measured from hydrogen versus metal/He lines are not uncommon, and have been described in the literature by several authors (e.g., \citealt{bagn2006}). Generally, they are understood to be a result of the modification of the Stokes $V$ profiles due to nonuniform weighting of the flux contributions from different parts of the stellar surface caused by inhomogeneous abundance distributions of different chemical elements. However, in the case of HD 184927, the effect appears to be exceptionally strong, leading to more than a factor-of-two difference in the maximum longitudinal field.

To allow us to explore this phenomenon in more detail later in this paper, we have obtained additional measurements of the longitudinal field. First, we used Eq.~(\ref{bzeqn}) to measure \bz\ from various individual spectral lines of He and different metals. These lines are summarized, along with their relevant atomic data, in Table~\ref{tab:indlines}. Secondly, we used the LSD line mask described in Sect. \ref{sect:espadons} to generate new sub-masks including only lines of single chemical elements. Sufficient lines were available to develop such masks for the elements He -- 54 lines, N -- 63 lines, O -- 92 lines, Si -- 22 lines and Fe -- 57 lines. These masks were then used to extract new LSD profile sets, from which we measured the longitudinal field using Eq.~(\ref{bzeqn}). These results will be discussed later in the paper.

All of the longitudinal field measurements derived from LSD profiles are summarized in Table~\ref{tab:elems}. All of the measurements obtained from individual lines are summarized in Table~\ref{tab:lines}.


\begin{table}
\caption{Summary of wavelengths and Land\'e factors of individual lines analyzed for longitudinal magnetic fields.} 
\begin{tabular}{lcr}
\hline
\hline
Ion & Wavelength & Land\'e  \\
& (\AA) &factor\\
\hline
H  {\sc i} & 6562.801 & 1.0\\
H  {\sc i} & 4861.363 & 1.0\\
He  {\sc i} & 6678.154 & 1.0 \\
Fe  {\sc ii} & 5156.111&1.25\\
Si  {\sc iii} & 4552.622&1.25\\
C  {\sc ii} & 6578.052& 1.167\\
C  {\sc ii} & 6582.881&1.333\\
N  {\sc ii} & 4041.310&1.5\\
O  {\sc i} & 7771.942&1.5\\
O  {\sc i} & 7774.161&1.2\\
O  {\sc i} & 7775.388&1.5\\
\hline
\end{tabular}
\label{tab:indlines}
\end{table}

\section{Period of rotation}

The rotation period of HD\,184927 is well known, but our new magnetic field and line strength measurements enable us to improve on the period of $9.^{\!\!\rm{d}}52961 \pm 0.^{\!\!\rm{d}}00731$ established by \cite{wade97}. Making the assumption that to first order the magnetic field of the star is approximately dipolar, we have determined periods by fitting  various subgroups of magnetic field measurements to the equation $B_e=B_0+B_1\sin{2\pi(\phi-\phi_0)}$. First, because of their very high S/N ratio, the LSD \bz\,values in the first column of Table \ref{tab:elems} alone provide a better period determination of $9.^{\!\!\rm{d}}53089 \pm 0.^{\!\!\rm{d}}00200$;  the reduced $\chi^2$ of the best fit sinusoid is 0.93. Secondly, a periodogram analysis of the combined magnetic data from \cite{wade97} and the new $dimaPol$ H$\beta$ measurements from Table \ref{dima:data} yield a still more accurate period of $9.^{\!\!\rm{d}}53071 \pm 0.^{\!\!\rm{d}}00120$, albeit with a higher reduced $\chi^2$ of 2.24.  

The $dimaPol$ He~{\sc i} field measurements show a clear offset from the H$\beta$ and earlier published data and so were not included in the analysis above.  There is also a very pronounced difference between the amplitude of the \bz\, field curve produced from our LSD analysis of the ESPaDOnS data using metal and helium lines and that derived from the various H$\beta$ measurements as well as H$\alpha$ line measurements. Because of this we have also performed a period analysis for only the combined H$\beta$ measurements of \cite{wade97}, the $dimaPol$ H$\beta$ measurements, and the ESPaDOnS magnetic field values derived from the H$\alpha$ line given in the last column of Table \ref{tab:elems}.  These data combined give only a single possible period: 
$9.^{\!\!\rm{d}}53083 \pm 0.^{\!\!\rm{d}}00090$ ($\chi^2 = 2.20$).  

We can further improve the period by using the very long time baseline between the He~{\sc i} line intensity measurements of \cite{bond76} and measurements from our new ESPaDOnS data discussed in the next section. We assume that the period has not changed significantly during the last 40 years. One can test this assumption by evaluating the agreement of a constant-period model with the data. A simple visual inspection of the relative phasing of the two line strength curves (with the \cite{bond76} data arbitrarily scaled) yields a final best-fit period of $9.^{\!\!\rm{d}}53102 \pm 0.^{\!\!\rm{d}}0007$. This is a 10-fold improvement in the uncertainty in the period provided by \cite{wade97}. If we adopt the zero point of the phase curve as the positive extremum of the magnetic field as determined by the best sinusoidal fit to the H$\beta$ field curve with this rotation period, we can then establish a new ephemeris for HD\,184927 of

\begin{equation}
\rm{JD}(B_e^+)=(2455706.517\pm0.48) + (9.53102\pm0.0007) \cdot \rm{E}\label{ephem}
\end{equation}

\noindent with values of $965 \pm 7$\,G and  $-889 \pm 13$\,G for $B_0$ and $B_1$ respectively.

\begin{figure}
\includegraphics[scale=0.34]{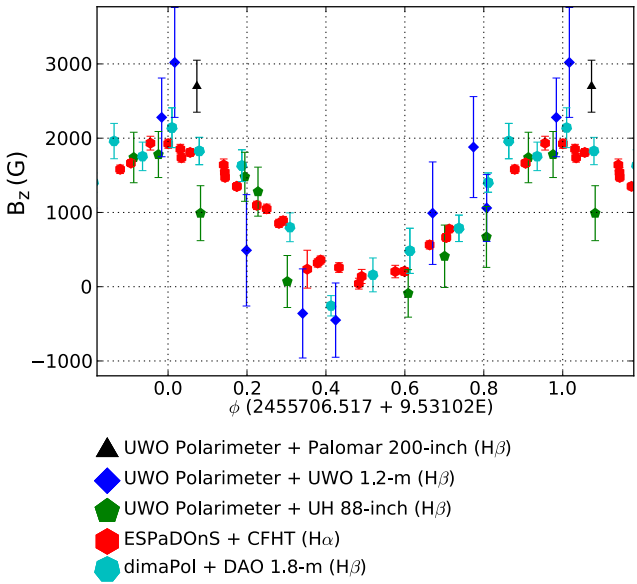}
\caption{Variation of longitudinal field measured using H lines, phased using the new ephemeris.}
\label{fig:period}
\end{figure}

The magnetic field curve of HD\,184927 obtained from H lines is shown in Figure \ref{fig:period}. The helium line strength maximum (Fig.~\ref{fig:helium}) occurs very close to the positive magnetic field extremum.

\begin{table*}
\caption{Magnetic field measurements using LSD masks.}
\begin{tabular}{ccrrrrrr}
\hline
HJD  & Phase  & Full &   Fe&  Si&O& N & He \\
&&(G)&(G)&(G)&(G)&(G)&(G)\\
\hline
2454667.9326&0.031& 1017 $\pm$ 19&  822 $\pm$ 115& 1265 $\pm$ 28&  965 $\pm$ 76& 1034 $\pm$ 52& 1028 $\pm$ 36\\
2454667.9593&0.034& 1021 $\pm$ 18&  674 $\pm$  98& 1239 $\pm$ 26& 1016 $\pm$ 79&  998 $\pm$ 51& 1038 $\pm$ 36\\
2454698.9072&0.281&  298 $\pm$ 12& -300 $\pm$  53&   87 $\pm$ 16&  -69 $\pm$ 28& -115 $\pm$ 16&  526 $\pm$ 25\\
2454699.8416&0.379&  -30 $\pm$ 11& -638 $\pm$  68& -318 $\pm$ 15& -467 $\pm$ 36& -494 $\pm$ 19&  258 $\pm$ 22\\
2454699.9123&0.386&  -42 $\pm$ 11& -643 $\pm$  67& -365 $\pm$ 15& -496 $\pm$ 37& -486 $\pm$ 20&  243 $\pm$ 22\\
2454700.8352&0.483& -216 $\pm$ 14& -805 $\pm$  75& -585 $\pm$ 21& -681 $\pm$ 48& -703 $\pm$ 26&   79 $\pm$ 27\\
2454700.9051&0.491& -214 $\pm$ 18& -763 $\pm$  82& -603 $\pm$ 27& -692 $\pm$ 56& -733 $\pm$ 32&   71 $\pm$ 33\\
2454754.7586&0.141&  817 $\pm$ 22&  498 $\pm$  92&  889 $\pm$ 30&  718 $\pm$ 66&  596 $\pm$ 39&  884 $\pm$ 38\\
2454754.7858&0.144&  810 $\pm$ 17&  345 $\pm$  71&  925 $\pm$ 23&  670 $\pm$ 56&  618 $\pm$ 34&  937 $\pm$ 31\\
2454756.7762&0.353&  126 $\pm$ 52& -642 $\pm$ 213& -412 $\pm$ 90& -412 $\pm$170& -443 $\pm$ 70&  277 $\pm$108\\
2454759.7313&0.663&   15 $\pm$ 13& -607 $\pm$  62& -473 $\pm$ 17& -495 $\pm$ 38& -655 $\pm$ 24&  330 $\pm$ 25\\
2454954.0990&0.056& 1003 $\pm$ 16&  607 $\pm$  88& 1257 $\pm$ 23&  886 $\pm$ 68&  966 $\pm$ 48& 1053 $\pm$ 28\\
2454954.9434&0.144&  812 $\pm$ 15&  361 $\pm$  64&  904 $\pm$ 19&  641 $\pm$ 52&  625 $\pm$ 32&  923 $\pm$ 27\\
2454955.9478&0.250&  424 $\pm$ 15& -130 $\pm$  53&  327 $\pm$ 22&   95 $\pm$ 34&   12 $\pm$ 19&  644 $\pm$ 28\\
2454959.0524&0.576& -192 $\pm$ 15& -836 $\pm$  79& -633 $\pm$ 23& -728 $\pm$ 53& -726 $\pm$ 28&  133 $\pm$ 30\\
2454963.1008&0.000& 1027 $\pm$ 17&  657 $\pm$  93& 1228 $\pm$ 25&  998 $\pm$ 75&  942 $\pm$ 46& 1077 $\pm$ 31\\
2455019.1256&0.879&  828 $\pm$ 16&  317 $\pm$  55&  812 $\pm$ 23&  576 $\pm$ 47&  554 $\pm$ 31&  941 $\pm$ 28\\
2455251.1717&0.225&  503 $\pm$ 14&   -8 $\pm$  48&  498 $\pm$ 20&  214 $\pm$ 33&  196 $\pm$ 22&  692 $\pm$ 28\\
2455253.1548&0.433& -161 $\pm$ 12& -687 $\pm$  68& -512 $\pm$ 18& -588 $\pm$ 42& -652 $\pm$ 24&  145 $\pm$ 24\\
2455258.1328&0.955&  992 $\pm$ 23&  631 $\pm$  99& 1083 $\pm$ 35&  913 $\pm$ 77&  910 $\pm$ 50& 1086 $\pm$ 41\\
2455394.7707&0.291&  263 $\pm$ 13& -327 $\pm$  56&    1 $\pm$ 19& -165 $\pm$ 31& -186 $\pm$ 17&  517 $\pm$ 26\\
2455522.6867&0.712&  206 $\pm$ 14& -454 $\pm$  55& -231 $\pm$ 17& -304 $\pm$ 32& -508 $\pm$ 24&  475 $\pm$ 26\\
2455879.7358&0.174&  727 $\pm$ 15&  214 $\pm$  54&  720 $\pm$ 19&  455 $\pm$ 40&  475 $\pm$ 27&  833 $\pm$ 27\\
2456102.9978&0.599& -142 $\pm$ 13& -749 $\pm$  68& -585 $\pm$ 19& -720 $\pm$ 47& -746 $\pm$ 26&  158 $\pm$ 24\\
2456104.0072&0.705&  160 $\pm$ 15& -487 $\pm$  60& -253 $\pm$ 21& -352 $\pm$ 37& -567 $\pm$ 27&  421 $\pm$ 26\\
2456105.9222&0.906&  933 $\pm$ 15&  492 $\pm$  71&  941 $\pm$ 24&  736 $\pm$ 55&  612 $\pm$ 32&  968 $\pm$ 26\\
\hline                                                       
\end{tabular} 
\label{tab:elems}
\end{table*}

\section{Variability}

HD~184927 is one of the most slowly-rotating He-strong stars, leading to sharp Stokes $I$ and $V$ profiles. In this section we examine the equivalent width (EW) and line profile variations of various chemical elements, in order to understand their surface abundance distributions and relationship to the magnetic field.

\subsection{Equivalent widths}

We have computed EW variations of various representative spectral lines present in the ESPaDOnS spectra. Lines were selected based on their strength, lack of blends and clean continuum and are representative of the elements He, C, N, O, Si, and Fe. Each spectral line was renormalised to the local continuum before measuring EW. To estimate uncertainties in EW we used the prescription of \cite{vollmann}.

\begin{figure*}
\includegraphics[scale=0.44]{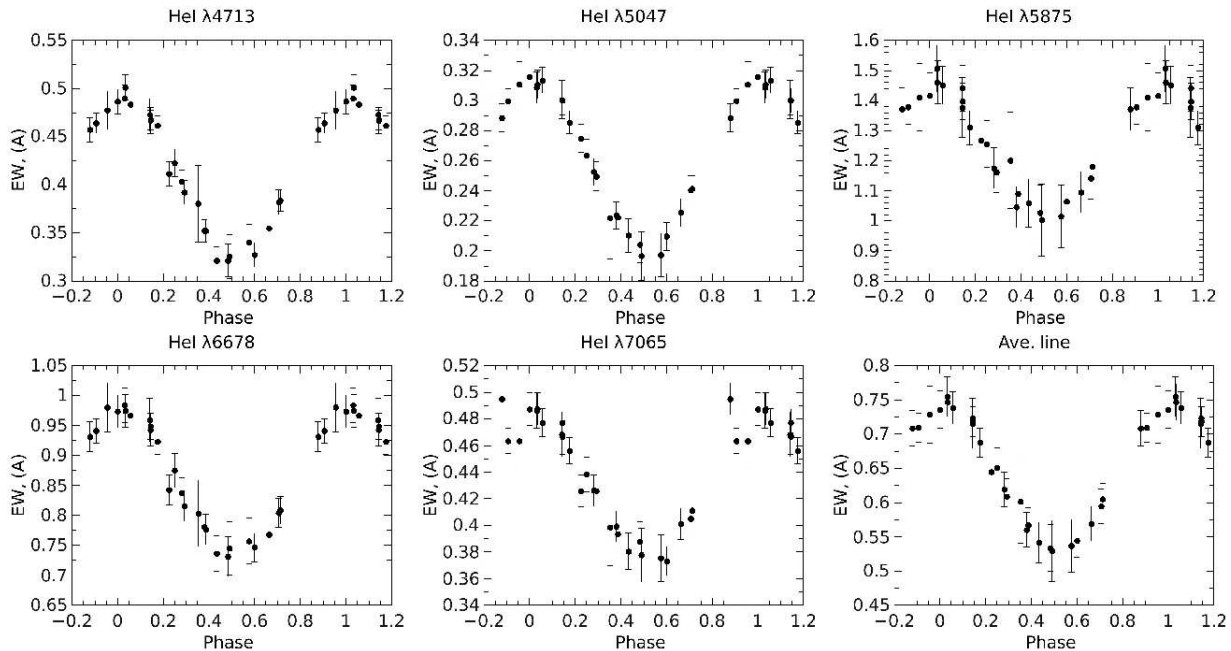}
 \caption{EW variations of helium lines in the spectrum of HD 184927. The mean variation at lower right is obtained by normalising each curve to its maximum and averaging at each phase.}
 \label{fig:helium}
\end{figure*}

At first we investigated the variability of helium spectral lines. Results for the lines He~{\sc i}~$\lambda\lambda$4713, 5047, 5876, 6678 and 7065 are shown in Fig. \ref{fig:helium}. Many He lines have broad wings that are blended with narrow lines of other elements; this fact complicates the measurement. We have chosen five lines which are least affected by blending. All the lines have similar sinusoidal variations when phased with the adopted rotation period. Absorption of all lines reaches its minimum at phase $\sim0.45-0.5$ and maximum near phase 0.0. At the bottom right of Fig. \ref{fig:helium} we present the mean variation computed from the normalized and averaged EWs of the five individual lines.

\begin{figure*}
\includegraphics[scale=0.37]{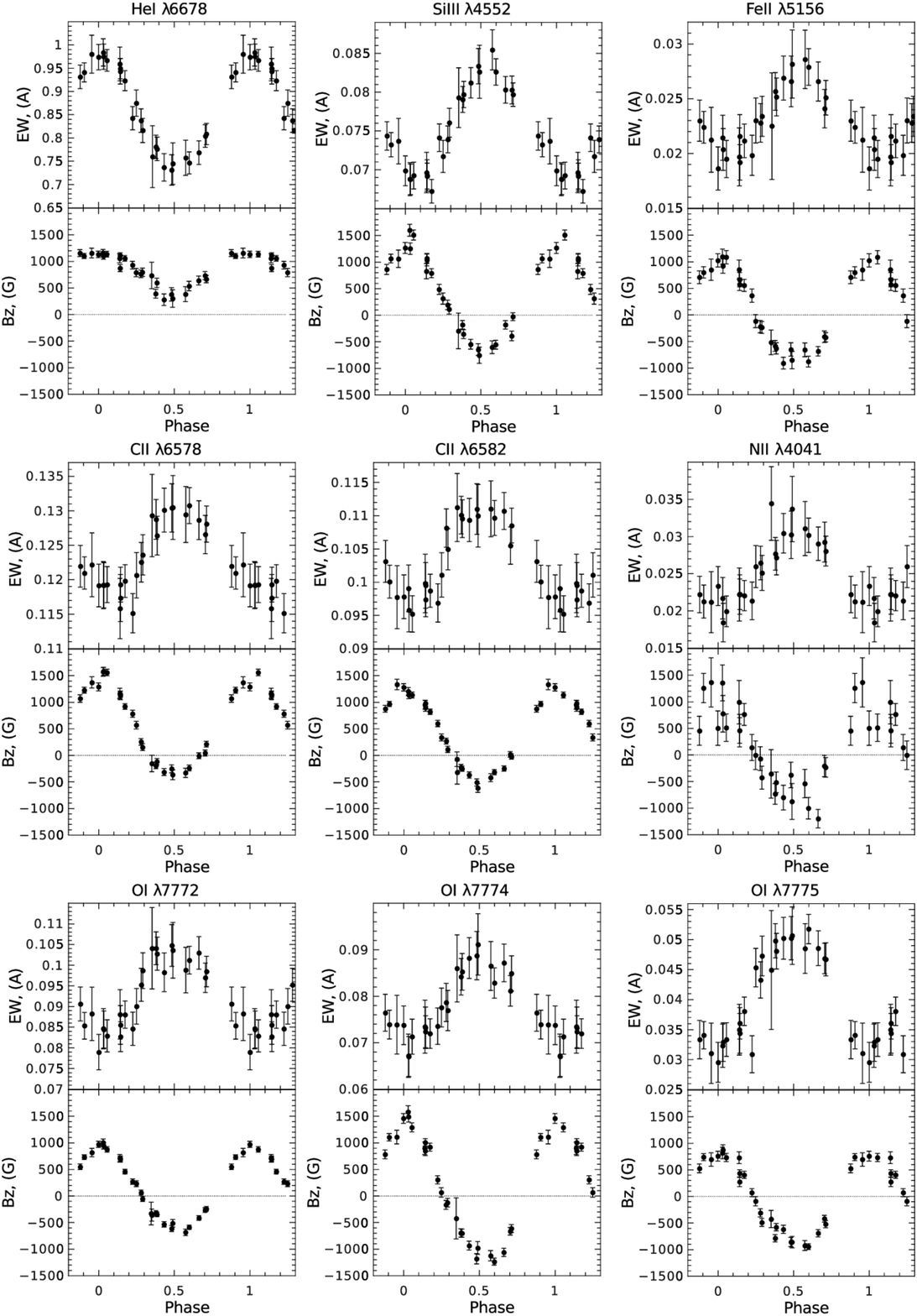}
 \caption{Phased EW measurements (top panels) and longitudinal magnetic field measurements (bottom panels) for selected spectral lines. Note that while maximum absorption of helium corresponds to magnetic field minimum, EWs of other lines varies opposite way.}
 \label{fig:ewbz}
\end{figure*}

In addition to helium we investigated the EW variations of a number of other spectral lines corresponding to different elements. We selected strong, unblended lines of the main ions presented in our spectra. Illustrative results are shown in Fig.~\ref{fig:ewbz}, along with the longitudinal magnetic field measured for each line (described in Sect.~\ref{sect:espmag}). The EWs of lines of carbon, oxygen, iron, nitrogen and silicon all vary sinusoidally, in a sense opposite to that of helium (i.e. with EW maxima at phase 0.5, and minima at phase 0.0).

Our measurements of EWs therefore suggest that HD 184927 exhibits a rather limited variety of line profile variability.

\begin{figure*}
\includegraphics[width=17cm]{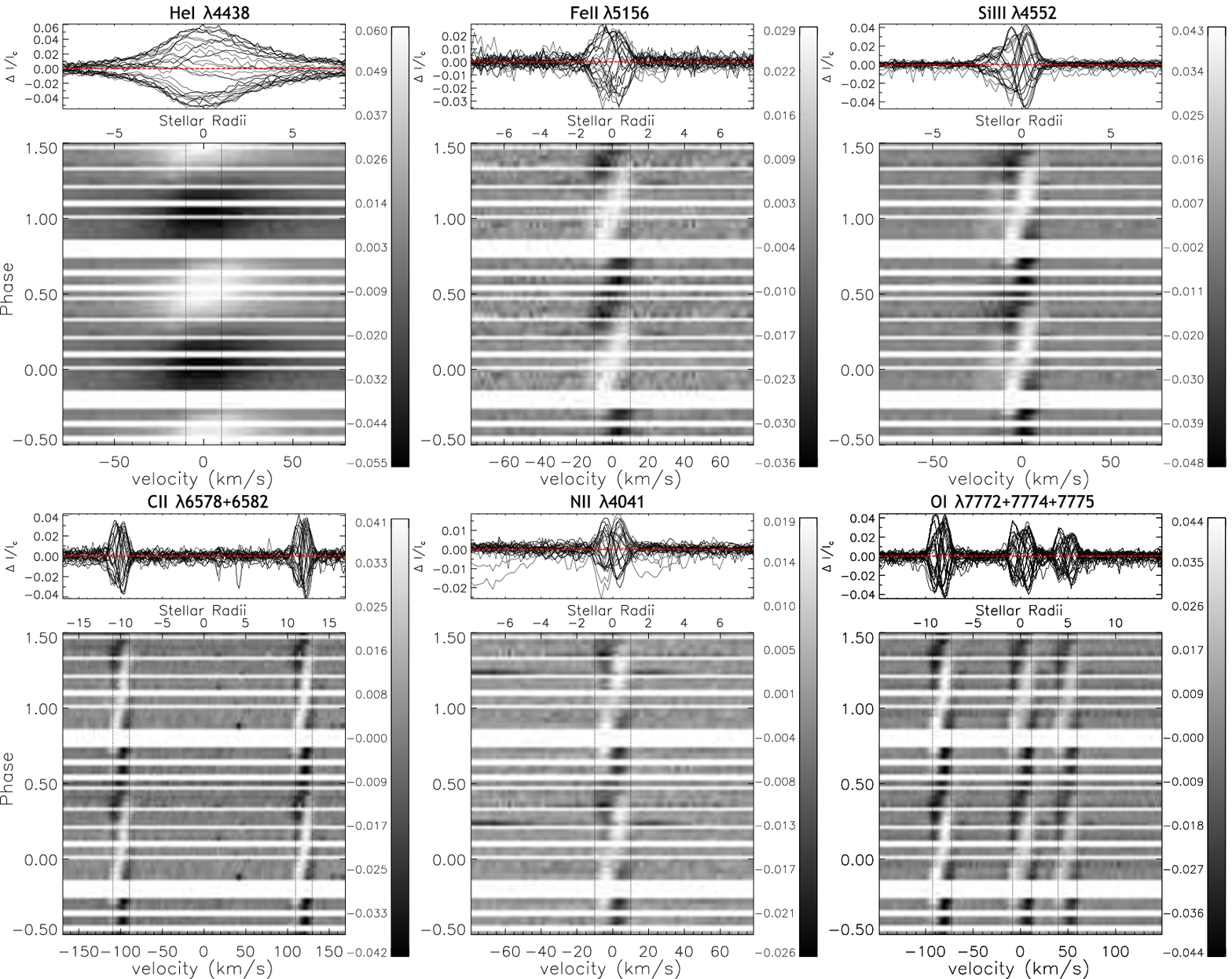}
 \caption{Dynamic spectra of selected lines. Vertical dashed lines correspond to $\pm v\sin i$. Upper panels display differences related to the mean line.}
 \label{fig:dynsp}
\end{figure*}

In Fig. \ref{fig:dynsp} we illustrate dynamic spectra of representative spectral lines of HD 184927. In these plots, individual spectra are presented as horizontal bands stacked vertically according to phase. Intensity, relative to the mean profile averaged over all phases, is identified with different colours. Systematic radial velocity~(-13 \kms) has been subtracted.

All the elements, except helium, show a similar picture on which pseudoabsorptions and pseudoemissions are travelling from negative to positive velocities. (The prefix ``pseudo-'' implies that emission and absorption features only appear relative to mean profile.) Usually these features represent regions in which the element is found to be overabundant (absorption) or under-abundant (emission) with respect to mean distribution in the atmosphere.

Emission features of all elements cross the point of null radial velocity $v=0$ \kms\, around phase 0.0 while helium is in strong absorption at that time. Later, the emission of metallic and CNO lines changes to absorption which crosses $v=0$ \kms\, at phase 0.5, when helium shows relatively weak emission. The line profile variations of all elements except helium (due to variability of the broad, Stark-broadened wings of lines of this element) occur inside the range of $\pm 10$ \kms\, (vertical dashed lines on Fig. \ref{fig:dynsp}), that is consistent with the $v\sin i$ parameter determined in Sect.~\ref{sect:atmos}. Thus the surface distribution of helium appears to be roughly opposite to those of the other elements of this star. In particular, behaviour of the helium lines suggest the presence of a large region of He overabundance visible on the stellar surface near phase 0.0.

\subsection{Longitudinal magnetic fields of individual elements and spectral lines}

As described in Sect.~\ref{sect:espmag}, we measured the longitudinal magnetic field using individual and combinations (via LSD) of spectral lines of different chemical elements, in order to explore the dependence of the field variation on chemical element.

\begin{figure}
 \center{\includegraphics[scale=0.27]{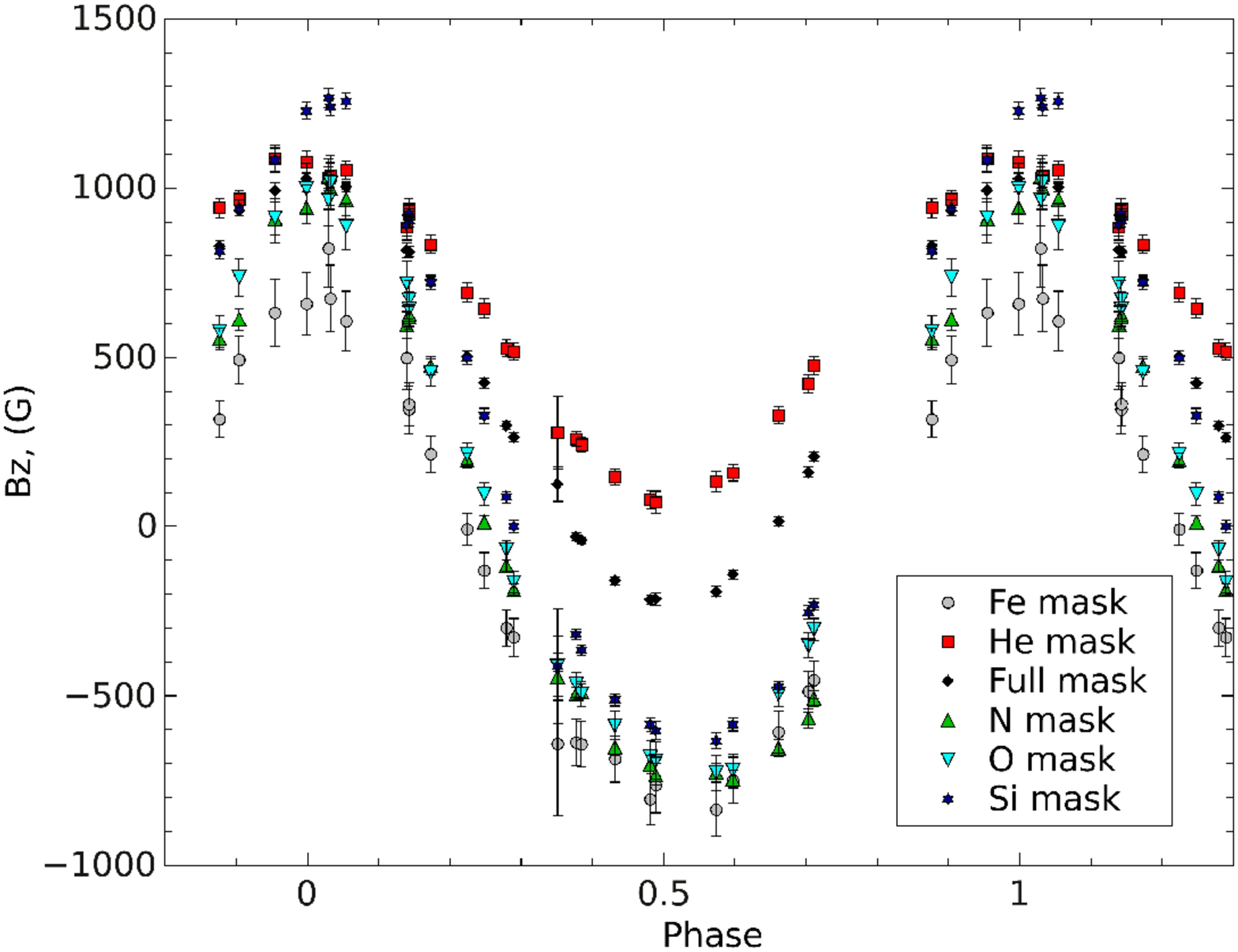}}
 \caption{Comparison of mean longitudinal field variations computed from LSD profiles which were generated using masks containing individual elements.}
 \label{fig:lsdelems}
\end{figure}

The phased longitudinal field data measured from the LSD profiles extracted for He, N, O, Si and Fe are illustrated in Fig. \ref{fig:lsdelems}, along with the variation obtained from the full mask (included for reference). One can see that the fields measured from N, O, Si and Fe vary approximately sinusoidally, and range from negative to positive values. While the negative extrema of these variations are mostly uniform (-600 to -800 G), the positive extrema vary from approximately +650 G for Fe to +950 G for N and O and +1200 G for Si. The variation obtained from the He mask is very different from those of the metals. In the case of the lower extremum, it is similar to that of the H lines. The lower extremum occurs at approximately zero, and consistent with the H lines the helium Stokes $V$ profiles show no inversion of their polarity. On the other hand, the positive extremum of the He field variation is much more similar to those of the metallic lines, with a peak value near 1 kG (i.e. approximately half that derived from the H lines). These results are roughly consistent with the dimaPol measurements from He~{\sc i} $\lambda 4922$.

The important systematic differences between the longitudinal field variations of these various elements are usually interpreted as due to different weightings of the flux contributions from different parts of the surface of the star as a result of different surface abundance distributions. Understanding these differences is important, because a naive interpretation of the different curves would lead to large systematic differences in the inferred field geometry. Ultimately, the field geometry inferred from the LSD profiles, either from one element or from the full mask, would be substantially incorrect. One subsidiary goal of this paper is to demonstrate that a unique and accurate field geometry can be derived for all elements by careful consideration of the role of both field and abundance distributions in the formation of line profiles.

Since ESPaDOnS spectra provide us with high signal-to-noise ratio profiles in both Stokes $V$ and $I$ parameters, we can also measure the longitudinal field from individual spectral lines of various elements, as described in Sect~\ref{sect:espmag}. This will allow a better evaluation of the significance of the differences between the longitudinal field curves of different elements. The results of these measurements are presented in the bottom frames of the subplots shown in Fig. \ref{fig:ewbz}. One can see that for all investigated lines the longitudinal field curves are sinusoidal with extrema at phases 0.0 and 0.5 when phased according to the new ephemeris. However, in agreement with the LSD measurements, the extreme values for lines of individual elements differ systematically. 

\begin{figure}
 \center{\includegraphics[scale=0.26]{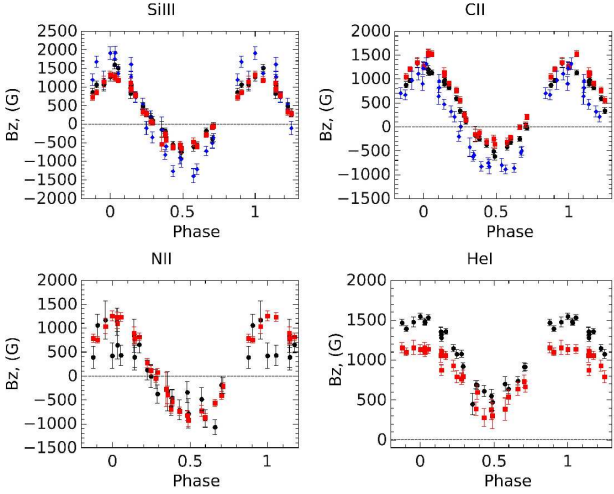}}
 \caption{Magnetic field from different lines of an element. Top left: black dots -- Si~{\sc iii}~$\lambda$\,4552, red squares -- Si{\sc iii}~$\lambda$\,4567, blue diamonds -- Si{\sc iii}~$\lambda$\,4574; top right: black dots -- C~{\sc ii}~$\lambda$\,6582, red squares -- C~{\sc ii}~$\lambda$\,6578, blue diamonds -- C~{\sc ii}~$\lambda$\,5145; bottom left: black dots -- N~{\sc ii}~$\lambda$\,4041, red squares -- N~{\sc ii}~$\lambda$\,5679; bottom right: black dots -- He~{\sc i}~$\lambda$\,4713, red squares -- He~{\sc i}~$\lambda$\,6678.}
 \label{fig:eldif}
\end{figure}

Nevertheless, as is apparent in Fig.~\ref{fig:ewbz}, the longitudinal field curves measured from different lines of the same element can sometimes also differ significantly. For example, for the 3 lines of the neutral O triplet at 777 nm, the longitudinal  field of the $\lambda7774$ line varies from -1.2 to 1.5 kG, whereas the remaining two lines of the triplet exhibit smaller variations: O~{\sc i} $\lambda7772$ varies from -680 to +1000 G, and O~{\sc i} $\lambda7775$ from -940 to 870 G. Comparable differences can be observed in the lines of some other elements. In Fig. \ref{fig:eldif} we illustrate the magnetic curves obtained from selected lines of Si, C, N and He. Some dispersion is again evident in each subplot.
This is likely explained by differential saturation of the lines. For example, a strong line would exhibit a smaller relative variation of its EW due to abundance spots in comparison to a weak line. Consequently, the longitudinal field derived from weak lines should be more significantly distorted by abundance inhomogeneities.  We examined the relative EW variations of the lines used to create this figure, and indeed found such a general relationship.

Notwithstanding these differences, considering that the apparent systematic differences are clearly enhanced when averaging many lines of the same element (Fig.~\ref{fig:lsdelems}), and that systematic differences are observed in the shapes of Stokes $V$ profiles observed at the extrema of the variation, we conclude that the observed systematic differences between the longitudinal field variations are real and significant.


The magnetic field measurements illustrated in Fig.~\ref{fig:eldif} are reported in Table~\ref{tab:lines}.

\begin{table*}
\addtolength{\tabcolsep}{-3pt}
\caption{Magnetic field measured from individual lines in gauss $\pm 1\sigma$ error bar. Phase according to  Eq. \ref{ephem}.}
\begin{tabular}{crrrrrrrrrrrrrrrrr}
\hline
Phase & Si{\sc iii} $\lambda4552$& Si{\sc iii} $\lambda4567$& Si{\sc iii} $\lambda4574$ &C{\sc ii} $\lambda5145$&C{\sc ii} $\lambda6578$&C{\sc ii} $\lambda6582$&N{\sc ii} $\lambda4041$& N{\sc ii} $\lambda5679$& He{\sc i} $\lambda4471$& He{\sc i} $\lambda6678$ & H$\alpha$ \\
&(G)&(G)&(G)&(G)&(G)&(G)&(G)&(G)&(G)&(G)&(G)\\
\hline
0.000& 1267 $\pm$ 103& 1328 $\pm$ 92& 1909   $\pm$ 174& 821  $\pm$ 102& 1262 $\pm$ 73& 1281 $\pm$ 69&423  $\pm$ 271& 1254 $\pm$ 106&1548 $\pm$ 40& 1132 $\pm$ 60   &$1664\pm  46$  \\
0.031& 1599  $\pm$ 112& 1295 $\pm$ 99& 1758   $\pm$ 201& 473  $\pm$ 102& 1537 $\pm$ 79& 1204 $\pm$ 73&1136 $\pm$ 283& 1227 $\pm$ 105&1482 $\pm$ 41& 1160 $\pm$ 64  &$1853\pm  64$  \\
0.034& 1253 $\pm$ 108& 1241 $\pm$ 95& 1917   $\pm$ 181& 323  $\pm$ 112& 1548 $\pm$ 75& 1136 $\pm$ 70&644  $\pm$ 285& 1097 $\pm$ 101&1473 $\pm$ 39& 1102 $\pm$ 61   &$1736\pm  59$  \\
0.056& 1509 $\pm$ 94& 1188  $\pm$ 80& 1368   $\pm$ 155& 211  $\pm$ 116& 1528 $\pm$ 61& 1136 $\pm$ 57& 432 $\pm$ 218& 1230 $\pm$ 91& 1532 $\pm$ 34& 1137 $\pm$ 54   &$ 853\pm  55$  \\
0.141& 829  $\pm$ 143& 1106 $\pm$ 122& 1096  $\pm$ 237& -198 $\pm$ 101& 1141 $\pm$ 96& 953  $\pm$ 85& 840 $\pm$ 346& 897  $\pm$ 130&1357 $\pm$ 54& 1057 $\pm$ 85   &$ 320\pm  56$  \\
0.144& 1039  $\pm$ 111& 951  $\pm$ 98& 1613   $\pm$ 184& 10   $\pm$ 99& 1105  $\pm$ 77& 898  $\pm$ 69& 389 $\pm$ 260& 770  $\pm$ 97& 1280 $\pm$ 42& 874  $\pm$ 66  &$ 359\pm  55$  \\
0.144& 1071 $\pm$ 91& 965   $\pm$ 83& 1277   $\pm$ 139& -424 $\pm$ 196& 1124 $\pm$ 57& 981  $\pm$ 54& 393 $\pm$ 206& 790  $\pm$ 83& 1324 $\pm$ 34& 1109 $\pm$ 54   &$  41\pm  73$  \\
0.174& 791  $\pm$ 84& 781   $\pm$ 80& 720    $\pm$ 132& -623 $\pm$ 89& 902   $\pm$ 53& 821  $\pm$ 50& 658 $\pm$ 178& 817  $\pm$ 82& 1360 $\pm$ 35& 1056 $\pm$ 53   &$ 138\pm  95$  \\
0.225& 487  $\pm$ 95& 330   $\pm$ 89& 551    $\pm$ 148& -589 $\pm$ 92& 768   $\pm$ 66& 600  $\pm$ 60& 122 $\pm$ 212& 272  $\pm$ 98& 1148 $\pm$ 45& 929  $\pm$ 72   &$1640\pm  80$  \\
0.250& 313   $\pm$ 106& 288  $\pm$ 102& -105  $\pm$ 174& -826 $\pm$ 99& 559   $\pm$ 71& 338  $\pm$ 65& -4 $\pm$ 236& 116  $\pm$ 100&1076 $\pm$ 53& 790  $\pm$ 77   &$1541\pm  63$  \\
0.281& 194  $\pm$ 89& 55    $\pm$ 80& 206    $\pm$ 151& -757 $\pm$ 112& 250  $\pm$ 61& 269  $\pm$ 56& -60 $\pm$ 194& -51  $\pm$ 82& 1078 $\pm$ 46& 765  $\pm$ 72   &$ 236\pm 255$  \\
0.291& 114  $\pm$ 92& 47    $\pm$ 78& -352   $\pm$ 148& -829 $\pm$ 148& 152  $\pm$ 63& 114  $\pm$ 59&-376 $\pm$ 190& 76   $\pm$ 81& 920  $\pm$ 48& 795  $\pm$ 78   &$ 563\pm  55$  \\
0.353& -142 $\pm$ 209& -537 $\pm$ 192& -153  $\pm$ 350& -799 $\pm$ 140& -149 $\pm$ 153&-182 $\pm$138&-321 $\pm$ 411& -394 $\pm$ 181& 448 $\pm$ 135& 448$\pm$ 135   &$1808\pm  53$  \\
0.379& -178 $\pm$ 79& -259  $\pm$ 74& -817   $\pm$ 133& -887 $\pm$ 91& -187  $\pm$ 58& -223 $\pm$ 53&-648 $\pm$ 172& -698 $\pm$ 75& 691  $\pm$ 54& 391  $\pm$ 83   &$1472\pm  50$  \\
0.386& -355 $\pm$ 84& -413  $\pm$ 75& -597   $\pm$ 133& -860 $\pm$ 90& -115  $\pm$ 56& -250 $\pm$ 51&-456 $\pm$ 172& -534 $\pm$ 74& 686  $\pm$ 56& 596  $\pm$ 88   &$1049\pm  63$  \\
0.433& -546 $\pm$ 91& -617  $\pm$ 87& -1258  $\pm$ 150& -536 $\pm$ 102& -307 $\pm$ 64& -370 $\pm$ 59&-710 $\pm$ 206& -734 $\pm$ 88& 611  $\pm$ 70& 279 $\pm$ 109   &$ 204\pm  84$  \\
0.483& -647  $\pm$ 111& -559 $\pm$ 99& -903   $\pm$ 175& -497 $\pm$ 90& -247  $\pm$ 74& -509 $\pm$ 67&-338 $\pm$ 216& -823 $\pm$ 99& 550  $\pm$ 81& 380 $\pm$ 126  &$1925\pm  61$  \\
0.491& -752 $\pm$ 152& -649 $\pm$ 132& -930  $\pm$ 225& 705  $\pm$ 118& -359 $\pm$ 90& -614 $\pm$ 82&-783 $\pm$ 302& -923 $\pm$ 136& 470 $\pm$ 106& 302$\pm$ 165   &$1581\pm  57$  \\
0.576& -602 $\pm$ 120& -477 $\pm$ 111& -1389 $\pm$ 192& 671  $\pm$ 102& -322 $\pm$ 85& -418 $\pm$ 75&-484 $\pm$ 241& -727 $\pm$ 116& 701 $\pm$ 92& 384 $\pm$ 140   &$1095\pm  56$  \\
0.599& -554 $\pm$ 83& -587  $\pm$ 74& -1213  $\pm$ 145& 973  $\pm$ 194& -231 $\pm$ 54& -316 $\pm$ 50&-892 $\pm$ 180& -861 $\pm$ 77& 636  $\pm$ 58& 537  $\pm$ 89   &$ 258\pm  65$  \\
0.663& -181 $\pm$ 79& -284  $\pm$ 72& -722   $\pm$ 127& 1112 $\pm$ 127& -7   $\pm$ 55& -247 $\pm$ 50&-1071$\pm$ 155& -567 $\pm$ 75& 738  $\pm$ 52& 640  $\pm$ 84   &$1933\pm  93$  \\
0.705& -388  $\pm$ 91& -73   $\pm$ 81& -501   $\pm$ 145& 903  $\pm$ 141& 44   $\pm$ 56& 20   $\pm$ 53&-186 $\pm$ 179& -407 $\pm$ 83& 915  $\pm$ 52& 732  $\pm$ 77  &$ 888\pm  58$  \\
0.712& -27  $\pm$ 74& -51   $\pm$ 68& -361   $\pm$ 115& 1227 $\pm$ 136& 207  $\pm$ 54& -21  $\pm$ 49&-204 $\pm$ 162& -205 $\pm$ 68& 914  $\pm$ 43& 667  $\pm$ 69   &$ 776\pm  50$  \\
0.879& 862  $\pm$ 95& 729   $\pm$ 85& 1197   $\pm$ 154& 1318 $\pm$ 120& 1048 $\pm$ 69& 877  $\pm$ 64& 390 $\pm$ 233& 790  $\pm$ 86& 1469 $\pm$ 41& 1156 $\pm$ 64   &$1351\pm  47$  \\
0.906& 1067 $\pm$ 82& 859   $\pm$ 70& 1679   $\pm$ 144& 642  $\pm$ 169& 1202 $\pm$ 54& 967  $\pm$ 48&1061 $\pm$ 232& 759  $\pm$ 72& 1393 $\pm$ 33& 1102 $\pm$ 48   &$ 207\pm  51$  \\
0.955& 1063  $\pm$ 164& 1142 $\pm$ 144& 1180  $\pm$ 252& 960  $\pm$ 130& 1345 $\pm$ 105&1334 $\pm$ 99&1168 $\pm$ 396& 1040 $\pm$ 159&1478 $\pm$ 62& 1157 $\pm$ 92  &$ 664\pm  52$  \\
\hline
\end{tabular}
\label{tab:lines}
\end{table*}


\section{Spectrum synthesis and fundamental parameters}
\subsection{Atmosphere}
\label{sect:atmos}

We used an extensive TLUSTY grid of atmosphere models to analyze the observed spectra of HD~184927. The TLUSTY code  \citep{hubeny95} incorporates the non-local thermodynamic equilibrium description (NLTE), 
including metal line-blanketing. It assumes a plane-parallel geometry and hydrostatic equilibrium. The physical parameters interval explored in the grid and the atomic data considered are appropriate for early B-type stars (see more details in \citealt{lanz07}).

\begin{figure*}
\includegraphics[scale=0.75, angle=90]{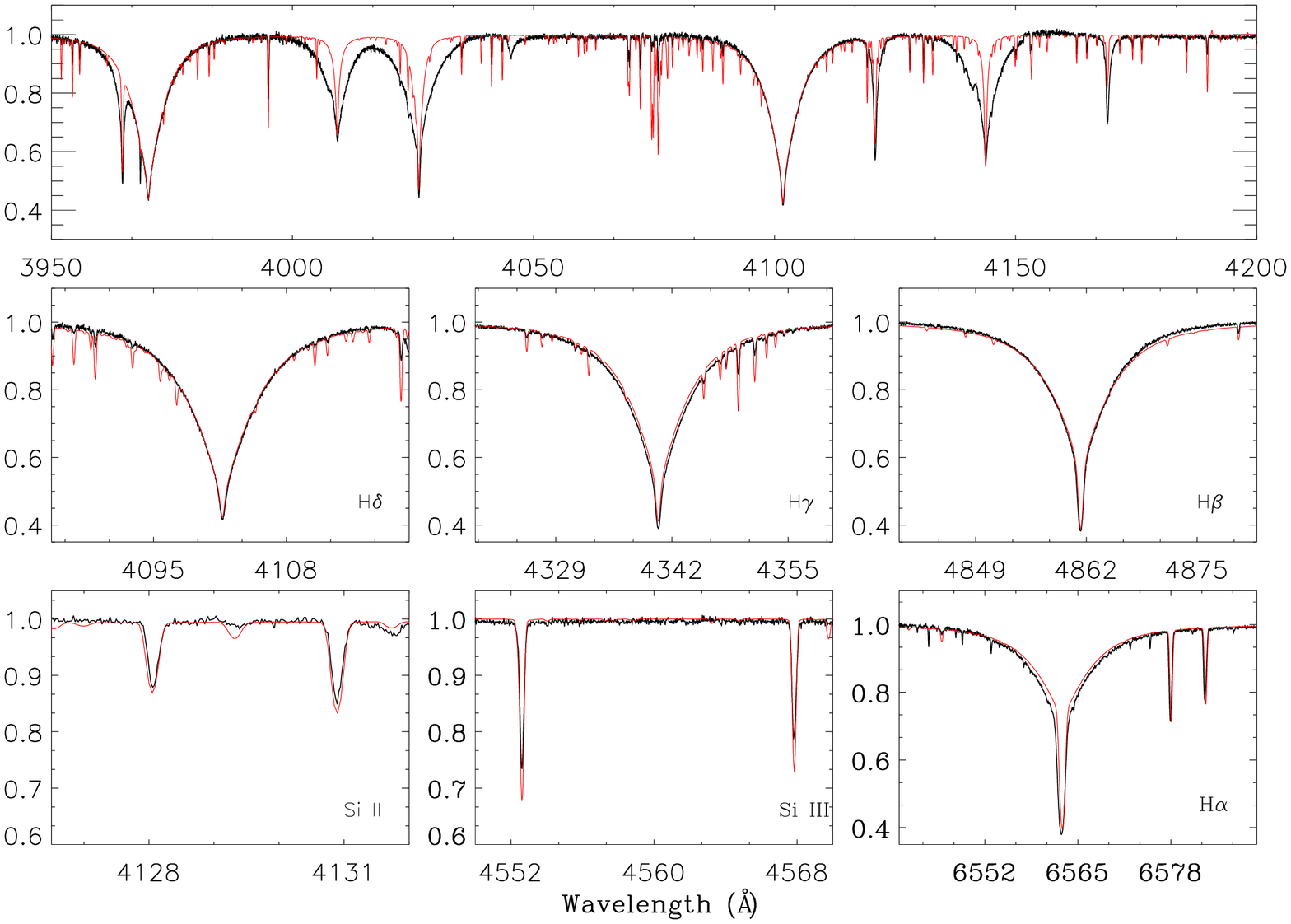}
\caption{TLUSTY model fit (red) to the spectrum of HD~184927 at He maximum (black). Note that the peculiar, intense helium profiles 
cannot be reproduced (see however Sect. 5.2). }
\label{fig:hemax}
\end{figure*}

\begin{figure*}
\includegraphics[scale=0.75, angle=90]{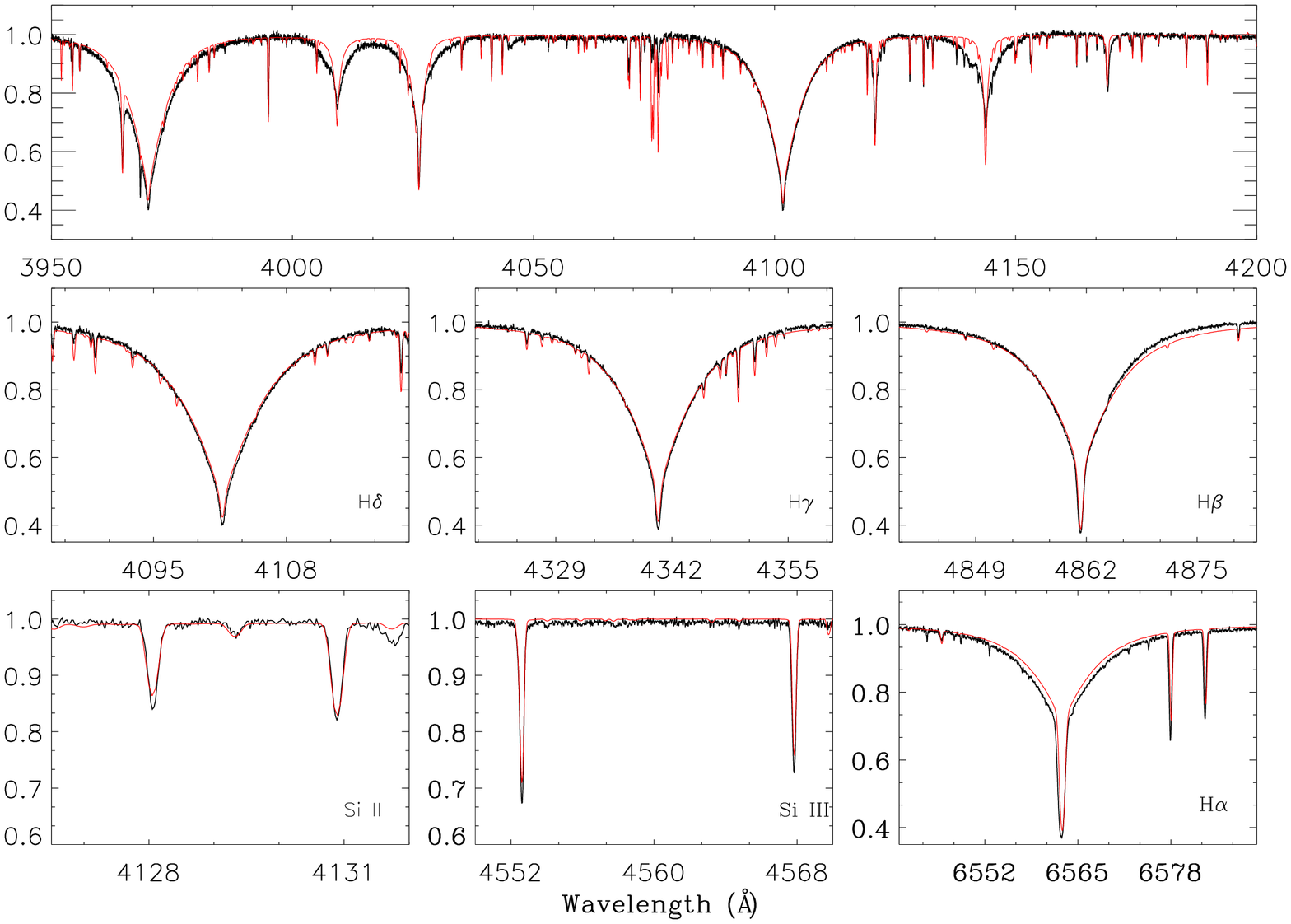}
\caption{TLUSTY model fit (red) to the spectrum of HD~184927 at He minimum (black). 
Note that a higher surface gravity compared to the case where He is at maximum was 
needed to fit the Balmer lines.}
\label{fig:hemin}
\end{figure*}

The spectrum of HD~184927 presents variability of various lines. To perform the analysis, we thus chose two (extreme) optical datasets: helium lines at maximum/minimum absorption (phases 0.000 and 0.491). Accordingly, we inferred a range of values for certain parameters (e.g. $\log g$). Our best fits are presented in Figs. \ref{fig:hemax} and \ref{fig:hemin} and the methodology used is described below.

The effective temperature ($\teff$) was derived from the balance of Si~{\sc ii-iii} lines (e.g. Si~{\sc ii} $\lambda$4124-31, Si~{\sc iii} $\lambda\lambda$4553, 4568). A simultaneous fit to these profiles for a specific temperature was only possible assuming a lower abundance of silicon (by $\sim$ 50\%), compared to the solar value \citep{grevesse09}. Both spectra (He max/min) could be matched with a $\teff$ of 22000 K. The surface gravity log $g$ was derived directly from the Balmer line wings. A log $g = 4.0$ (3.75) is determined for the helium minimum (maximum) absorption spectrum.

\begin{figure}
  \includegraphics[scale=0.28,angle=180]{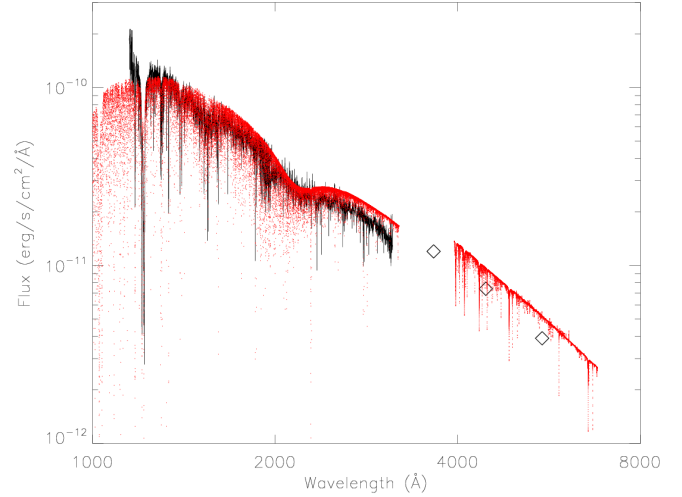}
  \caption{TLUSTY model (red) fit to the SED of HD 184927. IUE data (SWP+LWR) and UBV flux points are represented in black.}
  \label{fig:SED}
\end{figure}

The luminosity was estimated from fits to IUE data plus 
UBV photometry \citep{reed2003}. A simultaneous, perfect match for both spectral regions could not be obtained. The best fit to the observed continuum was achieved with 4000 L$_\odot$ 
and an E(B-V) = 0.15 (see Fig. \ref{fig:SED}), considering a compromise between the UV and optical
continuum level. The distance was kept fixed at 543 pc, 
corresponding to the revised Hipparcos parallax of HD~184927 
by \citeauthor{vanleeuwen07} (2007; $\pi = 1.84 \pm 0.55$). A luminosity of 
$\sim$10000 L$_\odot$, as previously inferred by \cite{wade97}, 
corresponds to a theoretical continuum much higher than the one 
observed.

The stellar radius was computed directly from the Stefan-Boltzmann equation: 
\begin{equation}
  R_\star = \left( \frac{L_\star}{4\pi\sigma T^4} \right)^{0.5}
\label{eqn:stb}
\end{equation}

\begin{figure}
\includegraphics[scale=0.41]{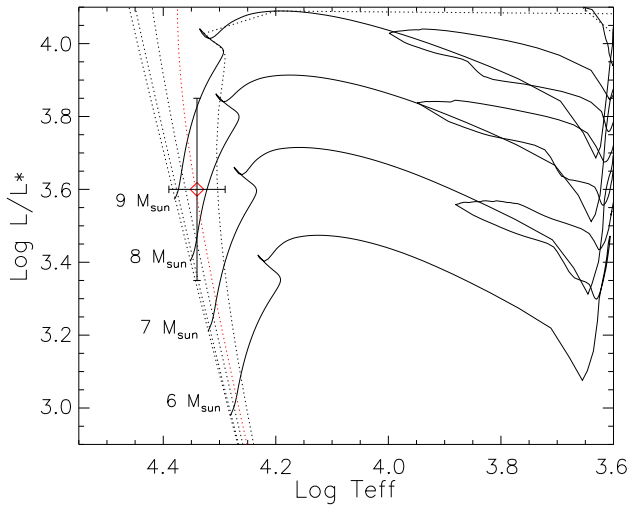}
  \caption{Position of HD 184927 in the H-R diagram. The evolutionary tracks (solid lines)
and isochrones (dotted lines) are from \citealt{georgy13}, appropriate for B stars.
The isochrones from the Main Sequence to later stages
are (log(age[yr])): 6.0, 6.5, 7.0, 7.25, and 7.5\,(left to right). The estimate age
for HD 184927 is 7.25 (about 18 Myr).}
  \label{fig:HRD}
\end{figure}

We placed HD 184927 on an HR diagram along with model evolutionary tracks from \cite{georgy13}, specific for B stars. We investigated tracks with and without rotation; ultimately, the differences are not significant. The position indicates a mass of $8.3\pm 0.7~M_\odot$ (see Fig. \ref{fig:HRD}). 

Using $\log g=4.0\pm 0.2$, $T_{\rm eff}=22000\pm 1000$~K (error bars were obtained from optical fits) and the formal error of the HIPPARCOS parallax we computed $\log L/L_\odot = 3.60^{+0.30}_{-0.22}$ and derived a radius $R/R_\odot = 4.4^{+1.9}_{-1.0}$. With this value we infer $M/M_\odot = 7.0^{+9.4}_{-3.5}$.  

If, on the other hand, we use the other $\log g=3.75$, we obtain $M/M_\odot = 4.0^{+5.4}_{-2.0}$. That is, in either case the spectroscopic mass is very uncertain, but in agreement with the evolutionary mass within the errors.

\begin{figure}
  \includegraphics[scale=0.37]{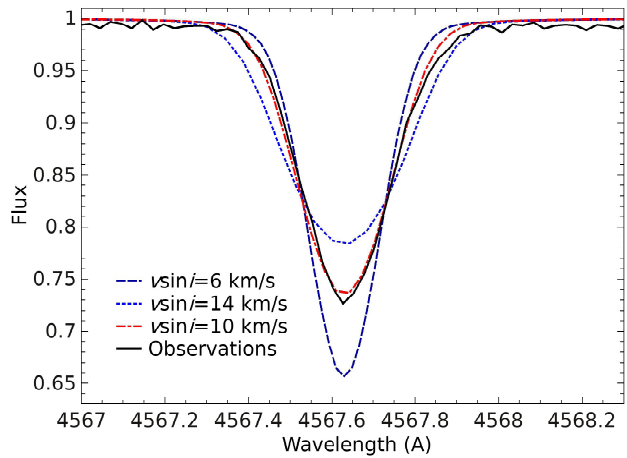}
  \caption{Comparison between the observed Si~{\sc ii} $\lambda 4567$ line (black) and models with different projected rotational velocities. The adopted $v\sin i=10\pm 2$~km/s.}
  \label{fig:vsini}
\end{figure}

In order to estimate projected rotational velocity we used a grid of NLTE TLUSTY models with fixed $\teff, \log g$, microturbulence and the combination of different [Si/H] and $v \sin i$ parameters to calculate several synthetic spectra for nine Si~{\sc ii-iii} lines. We used an atmosphere with fixed parameters $\teff=22000{\rm K}, \log g=4.0, V_{mic}=1 km s^{-1}$ and changed both [Si/H] from 1.0 to 0.3 (relative to solar abundance) and $v\sin i$ from 6 to 14 km\,s$^{-1}$. The phase 0.281 where Si field is about null was chosen. Then we used the code \verb|SYN_ABUND|\footnote{http://aegis.as.arizona.edu/~hubeny/pub/synplot2.1.tar.gz} to fit lines. It allows to fit two parameters (abundance of selected element and $v \sin i$). For each spectrum it computes the equivalent width and compares it to the EW determined from the observed spectrum. It also computes $\chi^2$ for each fit, and finds, for each line, a model that yields the lowest $\chi^2$, as well as that which yields the closest equivalent width to the observed one. Fig. \ref{fig:vsini} shows the best fit $v\sin i$ and $v \sin i \pm 4$ \kms for Si~{\sc iii} $\lambda$4567. For the following calculations we adopted a projected rotational velocity $v\sin i = 10 \pm 2$ km\,s$^{-1}$.

In this Section we have not estimated the impact of the strong magnetic field presented in HD~184927 on the formation of its spectrum. Magnetic splitting produced by the field can strongly affect the width of the line and lead to incorrect \vsini\ and turbulence determination, especially in the case of sharp lines (e.g. \citealt{neiner2012}). Unfortunately no NLTE magnetic synthesis code exists. \cite{prz11} showed that in fact LTE analysis can be used to determine parameters of stars up to $\teff=22000{\rm K}$ if lines are carefully selected. Note that HD~184927 has $\teff$ about 22000\,K, so it is difficult to judge which approach is more suitable in this case. In this paper we used full NLTE modeling to estimate the physical parameters of HD 184927 but in Section \ref{MDI}, dedicated to Magnetic Doppler Imaging, we make an attempt to estimate \vsini\ taking into account the presence of a magnetic field.

We obtain a good agreement between synthetic profiles and the main observed $\log g$ and $\teff$ diagnostics (i.e, Balmer and Si~{\sc ii-iii} transitions). However, we note that the helium lines present an important discrepancy: the observed wings are much broader than in the models \footnote{This fact confirms the status of HD~184927 as a He strong star (e.g., \citealt{higlee74})}. Moreover, we need an appropriate interpretation for the variations in the physical parameters needed to fit the different phases. This problem is addressed below, along with a possible solution based on an inhomogeneous atmosphere (i.e. lateral variations of the He abundance).

{\centering
\begin{table}
\caption{Summary of stellar parameters of HD~184927. Numbers in brackets correspond to the He maximum phase (see text in Sect. \ref{sect:atmos}).} 
\begin{tabular}{ll}                                          
\hline                                                       
Spectral type                 & B2V             \\             
$T_{\rm{eff}}$ (K)              & 22000 $\pm$ 1000 \\             
log $g$ (cgs)       & 4.00$\pm$0.2 (3.75)   \\                      
log L$_\star$/L$_\odot$         & 3.60 $^{+0.30}_{-0.22}$ \\
R$_{\star}$/R$_\odot$           & 4.4 $^{+1.9} _{-1.0}$  \\
M$_{\star}$/M$_{\odot}$  (evol)       &   $8.3\pm 0.7$  \\
M$_{\star}$/M$_{\odot}$  (spec)       & 7.0$^{+9.4}_{-3.5}$ (4.0$^{+5.4}_{-2.0}$)      \\
$v \sin i$ (km~s$^{-1}$)       &  10 $\pm$ 2     \\
\hline
\end{tabular}
\label{tab:params}
\end{table}
}
\subsection{Interpretation of He profiles with stratification and spots}

Because the helium line profiles of HD\,184927 at any particular rotational phase are not reproduced very well by LTE models with a uniform helium abundance and they also show a pronounced variation in strength, we have carried out a relatively simple NLTE investigation of some of these lines to see if a better fit to the profiles can be obtained with a patchy surface abundance of the element.


Again using the TLUSTY model atmosphere code we generated a grid of pure hydrogen and helium NLTE model atmospheres with $T_{\rm eff} = 22000$\,K, $\log g=4.0$ and 4.25 and a range of $N$(He/H) from 0.01 to 2.0.  The published line-blanketed TLUSTY model atmospheres BG22000g400v2 and BG22000g425v2 with $T_{\rm eff} = 22000$\,K, $\log g=4.0$ and 4.25 and $N$(He/H)$=0.1$ were used as the input starting atmospheres for the program. The companion spectral line synthesis program SYNSPEC was then used to produce corresponding line and continuum specific intensities for a spectral window containing the weak and strong He~{\sc i} $\lambda\lambda$4437 and 4471 lines for each helium abundance.

The specific intensities were then used in a disk integration program that permits the placement of circular zones of different helium abundances anywhere on the disk to define a simple helium surface abundance distribution. Each zone is defined by an angular radius, $R$, and its colatitude and longitude relative to a point that crosses the line of sight to the observer at $\phi = 0$.  
Bands of different abundances can also be modelled with the appropriate superposition of two circular patches with different radii. At each point of the disk integration the local helium abundance is determined from the spot locations and specific intensities for the appropriate local model atmosphere are used to construct the spectrum for that point.

Using the stellar radius, $v\sin i$ and period determined earlier in the paper, we employ the relation
\begin{equation}
 \frac{R}{R_{\odot}}=\frac{P v \sin i}{50.6 \sin i}
\label{eqn:sini}
 \end{equation}
 to compute a rotation axis inclination $i = 25\degr\pm 5\degr$. We adopt this value of $i$, and begin by assuming a magnetic obliquity of $\beta = 70\degr$. We generated spectra for $v\sin{i}$ between 0 and 20\,km\,s$^{-1}$. 

\begin{figure}
\includegraphics[scale=0.35]{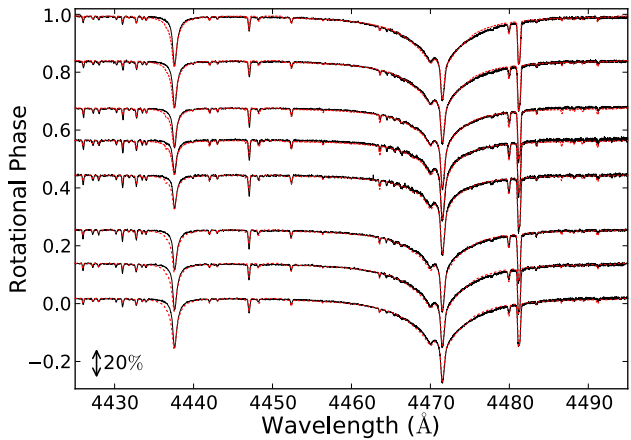}
\caption{Fits to He lines with the spotted model (red dashed line). See description in text.}
\label{fig:hespot}
\end{figure}

From the behaviour of the helium EW variations shown in Fig. \ref{fig:helium} it is not surprising that we find that the best model is one with a helium-rich spot centred at the positive magnetic pole of the star.  In fact, for our best model one entire hemisphere of HD\,184927 appears to have a very enhanced helium abundance with $N$(He/H)$\approx1.5$ while the remainder of the surface is helium-deficient with $N$(He/H)$\approx0.02$. 
Figure \ref{fig:hespot} shows a comparison of the synthetic spectra produced by this simple model to some of our ESPaDOnS spectra for a sample of rotational phases. A $v\sin{i}$ of 8\,km\,s$^{-1}$ (consistent with our determination from Si lines) provides the best fits to the metal lines in the spectral window.
We also note that the model spectra shown in the figure were produced with $\log{g}=4.25$; the wings of the He~{\sc i} $\lambda$4471 line are too weak with $\log{g}=4.0$ model atmospheres.

We have investigated similar models for different values of $i$ and $\beta$ and can reproduce the star's line profiles quite well for inclinations between 20 and 35$\degr$ and value of $\beta$ between 50 and $90\degr$.  The helium abundances in the enriched and deficient regions range from $N$(He/H)$=1.5$ to 2.0 and 0.01 to 0.05 respectively while the helium-rich 'patch' covers an angular area of between 140 to 200$\degr$ for the various model fits.  
This gives a sense of the uncertainties in the parameters for our simple NLTE surface abundance model.

These models support the values of $i$ derived above.  If the assumption that the star's field is approximately dipolar is valid, then $i + \beta$ must not be substantially larger than $90\degr$ since we do not see large negative values for \bz.  
We also conclude that the origin of the discrepancy of the helium line profiles in our earlier spectrum synthesis is a strong lateral variation of the abundance of the element across the stellar surface. In particular, we infer the presence of a large region strongly overabundant in helium that is visible near phase 0.0. This is consistent with our expectations based on the EW variation examined earlier in the paper.

\section {Magnetic Doppler Imaging}
\label{MDI}

\begin{figure*}
  \includegraphics[scale=0.8]{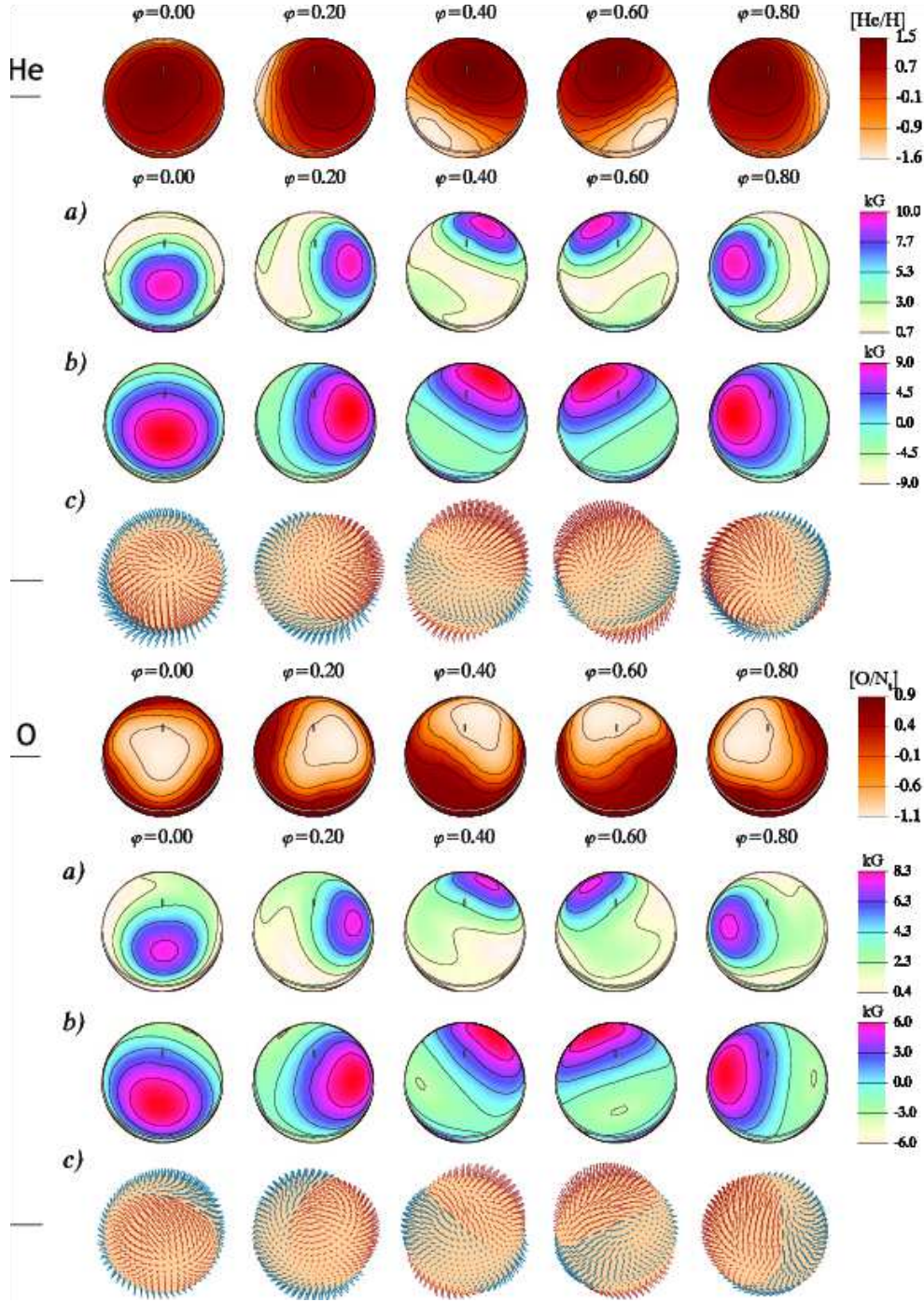}
  \caption{Results of Magnetic Doppler Imaging analysis of HD~184927. Top half: Chemical abundance and surface magnetic field distribution derived from the He~{\sc i} $\lambda$\,6678 line. Bottom half: The same for the O~{\sc i} $\lambda\lambda$\,7772, 7774, 7775 triplet. In each panel the top row shows spherical chemical abundance map, with scale given in logarithmic units relative to the Sun. The three rows below show a) field modulus, b) radial field component and c) field vector plot.}
  \label{fig:MDI}
\end{figure*}

Through the modelling carried out in the previous section, as well as line profile and EW variations, we infer that He has a strongly nonuniform distribution on the stellar surface, and other elements are distributed in a manner significantly different from He. We also observe important differences in the longitudinal magnetic field curves for different elements, which we speculate to be a consequence of these nonuniform distributions. To test these ideas and to derive a definitive magnetic field topology of HD\,184927, we employ the Magnetic Doppler Imaging method to map the magnetic field and abundance distributions for two elements, He and O. Our primary aim is to see if the magnetic field distributions derived independently from O and He lines are mutually compatible, and if a single field distribution is able to explain the diverse magnetic field curves that we observed. A secondary aim is to examine the distributions of these elements, and in particular compare to the parametric model developed in the last section for He. 

A non-solar He abundance and the presence of significant deviations from LTE in both He and O lines represent a major problem for the standard MDI approach \citep[e.g.][]{kp2002}, which assumes a fixed model atmosphere structure and fits the Stokes parameter spectra by varying local magnetic field parameters and individual abundances. Therefore, for this study we performed MDI inversions with a specially modified Invers13 code \citep{koch13}, originally developed for MDI of cool active stars. This code enables magnetic mapping of stellar surfaces with individual local atmospheres, fully accounting for the influence of the local abundance/temperature on the atmospheric structure, continuum fluxes, and absorption line profiles. For both He and O we computed grids of LLmodels atmospheres \citep{shulyak04} for $\teff=22000, \log g=4.0$ and a range of abundance values. For both elements the models were tabulated with a 0.25~dex step spanning a range from $-2.0$ to +1.5~dex relative to the solar abundances.

The polarised radiative transfer module of Invers13 was modified to account for the departures from LTE in both the source function and the line absorption coefficient according to the NLTE departure coefficients computed for LLmodels grids for the lines of interest. Only one chemical element at a time can be modelled in this way. Therefore, the modelling of He and O was carried out separately.

For the He~{\sc i} $\lambda$\,6678 line we performed NLTE calculations with TLUSTY, using a wrapper code developed by one of us (V. Tsymbal). These calculations employed the standard He model atom and other computational parameters as used for the B-star grid of \citet{lanz07}. For the oxygen triplet at $\lambda\lambda$ 7772, 7774 and 7775 we derived the NLTE departure coefficients using the following procedure. An extensive model atom including 51 levels of O~{\sc i} and the ground state of O~{\sc ii} was taken from \cite{przybilla00} and was updated as described by \citet{sit13} by implementing the electron collisional data from \cite{barklem07}. NLTE level populations were calculated using a revised version of the DETAIL code developed by \cite{butlergriddings85}. 

\cite{sit_iaus} showed that NLTE abundances determined from different lines of O~{\sc i} in reference A-B type stars give consistent NLTE abundances within the error bars, while the difference in LTE abundance from the IR (7771--5 \AA) lines and other O~{\sc i} lines from the visible spectral range was from 0.6 dex to 1.5 dex. 
 
The departures from LTE lead to strengthening of the O~{\sc i} IR lines, and the magnitude of the effect depends on oxygen abundance. In the case of a chemically uniform atmosphere, the NLTE abundance correction ($\Delta_{\rm NLTE} = \log\varepsilon_{\rm NLTE} - \log\varepsilon_{\rm LTE} $) for O~{\sc i} $\lambda$\,7771 line ranges between $-0.4$ dex and $-0.9$ dex.

\begin{figure*}
\centering
 \includegraphics[scale=0.7]{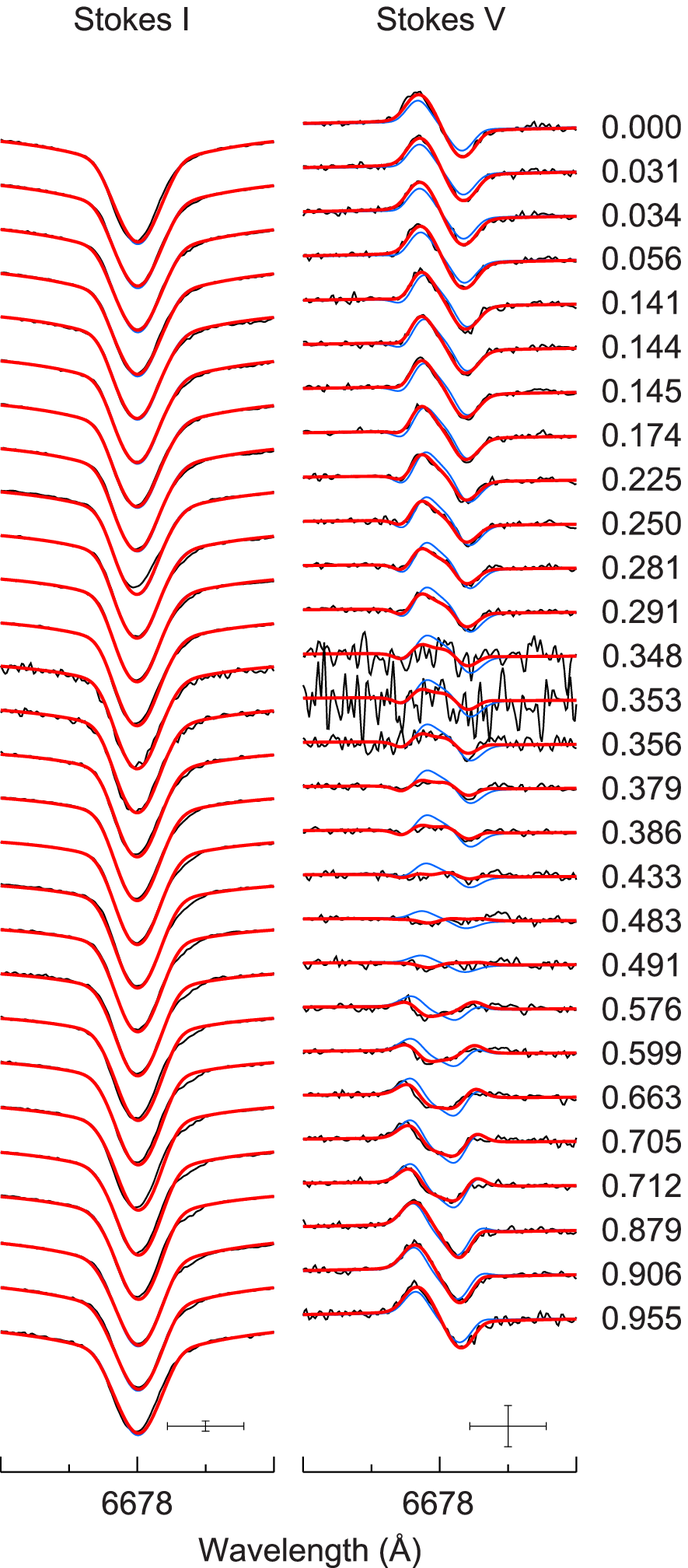}\hspace{1cm}\includegraphics[scale=0.7]{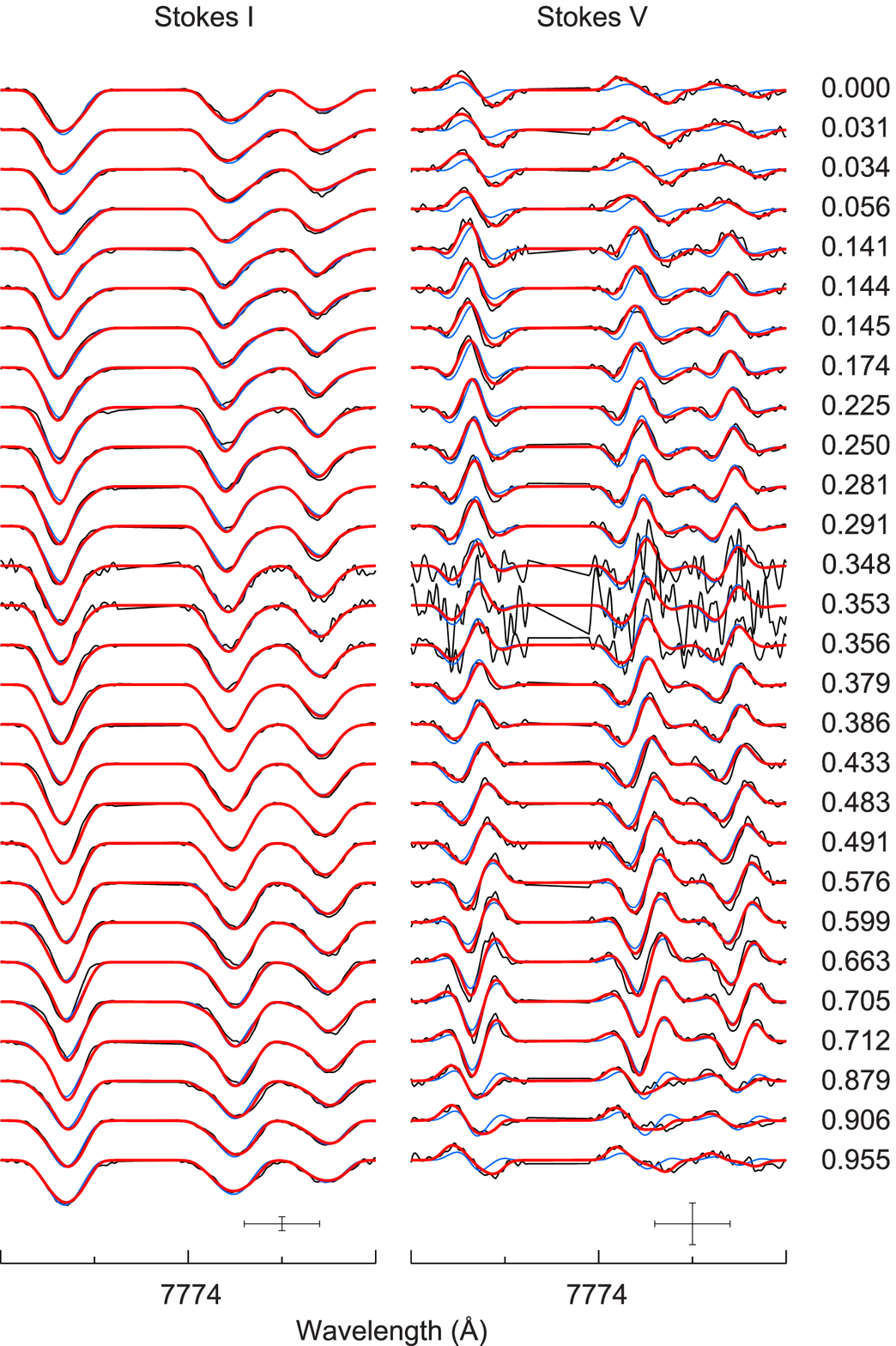}
 \caption{{\em Left -}\ Comparison between the observed (black lines) and synthetic (red lines) $I$ and $V$ Stokes spectra of the He~{\sc i} $\lambda$\,6678 line. {\rm Right -}\ Comparison between the observed and synthetic $I$ and $V$ Stokes spectra of O~{\sc i} $\lambda\lambda$\,7772, 7774, 7775. Blue line -- synthetic profiles computed under pure dipole assumption. Spectra for the consecutive rotational phases are shifted in the vertical direction. Rotational phases are indicated in the column to the right of the Stokes $V$ panel. The bars at the lower right of each panel show the horizontal (1 \AA) and vertical (5\%) scale.}
 \label{fig:mdi:prof}
\end{figure*}

All magnetic inversions adopted $i=25\degr$ and $v\sin i=10$~km\,s$^{-1}$. As discussed by \citet{koch13}, Invers13 can parameterise the surface magnetic field distribution with a spherical harmonic expansion, similar to the approach applied by \citet{donati06}. In this methodology the free magnetic mapping parameters are the spherical harmonic coefficients corresponding to the strength of the radial poloidal, horizontal poloidal and horizontal toroidal field for each pair of the angular degree $\ell$ and azimuthal order $m$. Given the low $v\sin i$ and unfavourable inclination, we do not expect to be able to map fine details of the surface magnetic field structure. Thus, we limited inversions to a purely poloidal dipolar configuration ($\ell_{\rm max}=1$) and a general quadrupolar field ($\ell_{\rm max}=2$), which included both poloidal and toroidal components. It turns out that only the latter parameterisation is capable of providing an adequate fit to the observed Stokes $IV$ profiles.

The top half of Fig.~\ref{fig:MDI} illustrates the abundance distribution of He, and the local orientation and strength of the magnetic field on the surface of HD~184927. The abundance map shows that almost all visible part of the star is overabundant with helium. In the same time, there is a large area near the equator which is best visible at phase 0.4--0.5 where the abundance of helium is about 3 times lower than on the remaining part of the surface. The zone of lack of helium lies in the area that is partly hidden due to low inclination, so we cannot predict the real scale of this area (note that we can see only $\sim60$\% of the stellar surface). Correlations between the surface magnetic field and abundance distribution of helium are pretty well seen. The area with the strongest surface magnetic field (roughly 9 kG) corresponds to the region where the abundance of helium is the highest.

The results of the MDI procedure applied to O~{\sc i} lines are presented in the bottom half of Fig.~\ref{fig:MDI}. The abundance map of oxygen features a large area of lower abundance of the element which rougly covers a zone coincident with the region of He overabundance. However, the spot where oxygen is lacking is less than the area of He overabundance. The other part of visible surface also shows deficit of the oxygen (relative to the Sun) but not so striking.



The MDI fits to the Stokes $I$ and $V$ profiles are illustrated in Fig.~\ref{fig:mdi:prof}. In addition to the adopted dipole + arbitrary quadrupolar field distribution (shown in Fig.~\ref{fig:MDI}), we also show the profiles computed by restricting the field distribution to a purely poloidal dipolar configuration. It is evident that the dipolar field fit does not reproduce all details of the Stokes $V$ profile variation. For instance, for the He~{\sc i} line the circular polarisation amplitude is overestimated in the phase interval 0.35--0.6; for the O~{\sc i} triplet the Stokes $V$ amplitude is somewhat underestimated around phase 0.05. We computed the reduced $\chi^2$ for Stokes V fits assuming pure dipole and dipole+quadrupole topologies. For He we obtained $\chi^2$=1.96\,for the dipole+quadrupole topology versus $\chi^2$=5.54\,for the pure dipole; for O we obtained $\chi^2$=3.70\,(dipole+quadrupole) and $\chi^2$=9.30\,(pure dipole). So, allowing only a dipolar field geometry increases the reduced $\chi^2$ by a factor of 2.5-5.  Therefore, we conclude that despite the sinusoidal appearance of the longitudinal field variations, the surface field topology of HD\,184927 includes a non-negligible quadrupolar component. At the same time, the field is predominantly poloidal. The relative magnetic energies of different harmonic terms are given in Table~\ref{tab:frac}. 

\begin{figure*}
 \includegraphics[width=16cm]{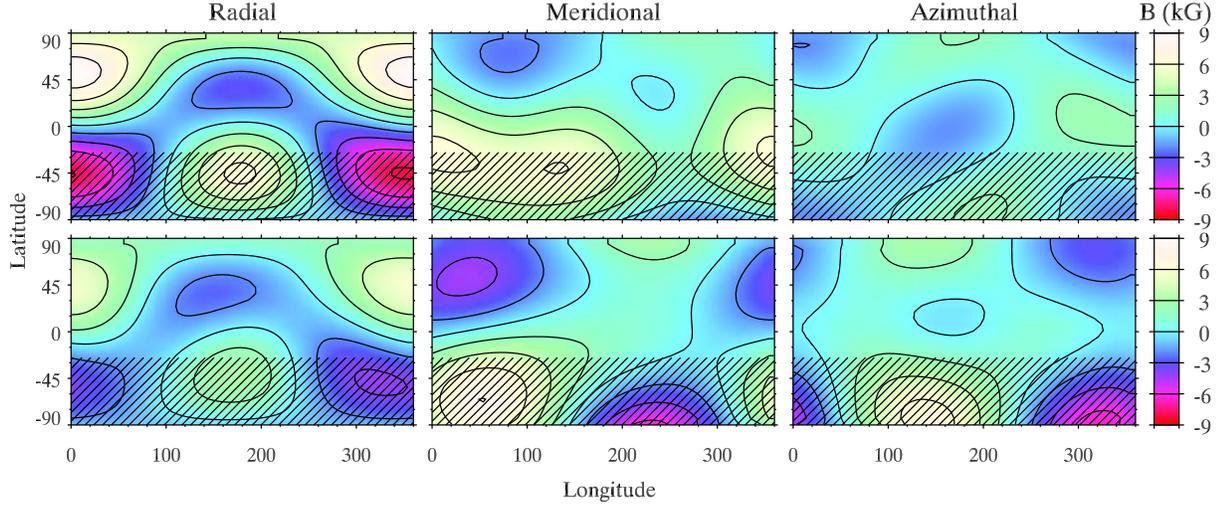}
 \caption{Comparison between the He magnetic map (upper plot) and O magnetic map (lower plot). The hatched area corresponds to the invisible part of the stellar surface.}
 \label{fig:mdi:com}
\end{figure*}

The two magnetic maps derived independently from He and O lines agree qualitatively. But some important local discrepancies are present. In particular, the amplitude of the field modulus is larger in the helium map (0.7--10~kG vs 0.4--8.3~kG for oxygen). In Fig.~\ref{fig:mdi:com} we compare rectangular plots of the three magnetic field components for the He (upper) and O (lower) magnetic maps. The rectangular maps are plotted using the same colour table and ranges. One can see that the He magnetic map is on average stronger. However, considering the rather extreme differences in the longitudinal field curves obtained for these two elements (see Fig.~\ref{fig:lsdelems}) and the stellar parameters that are not optimal for MDI, the agreement of the two magnetic maps is deemed to be satisfactory.

\begin{table}
\caption{Relative energies of different harmonic components of stellar magnetic field.} 
\begin{tabular}{cccc}                                          
\hline    
\hline    
 $\ell$ &  All   &  Poloidal   &  Toroidal\\
\hline    
 \multicolumn{4}{c}{He magnetic map} \\
  1 & 38.4\%  & 36.5\% &  1.9\% \\
  2 & 61.6\%  & 53.0\% &  8.5\% \\
All &        &   89.6\%& 10.4\% \\
 \multicolumn{4}{c}{O magnetic map} \\
  1 & 38.8\%  & 32.6\% &  6.2\% \\
  2 & 61.2\%  & 45.4\% &  15.8\% \\
All &        &   78.0\%& 22.0\% \\
\hline
\end{tabular}
\label{tab:frac}
\end{table}

\begin{figure*}
 \includegraphics[width=16cm]{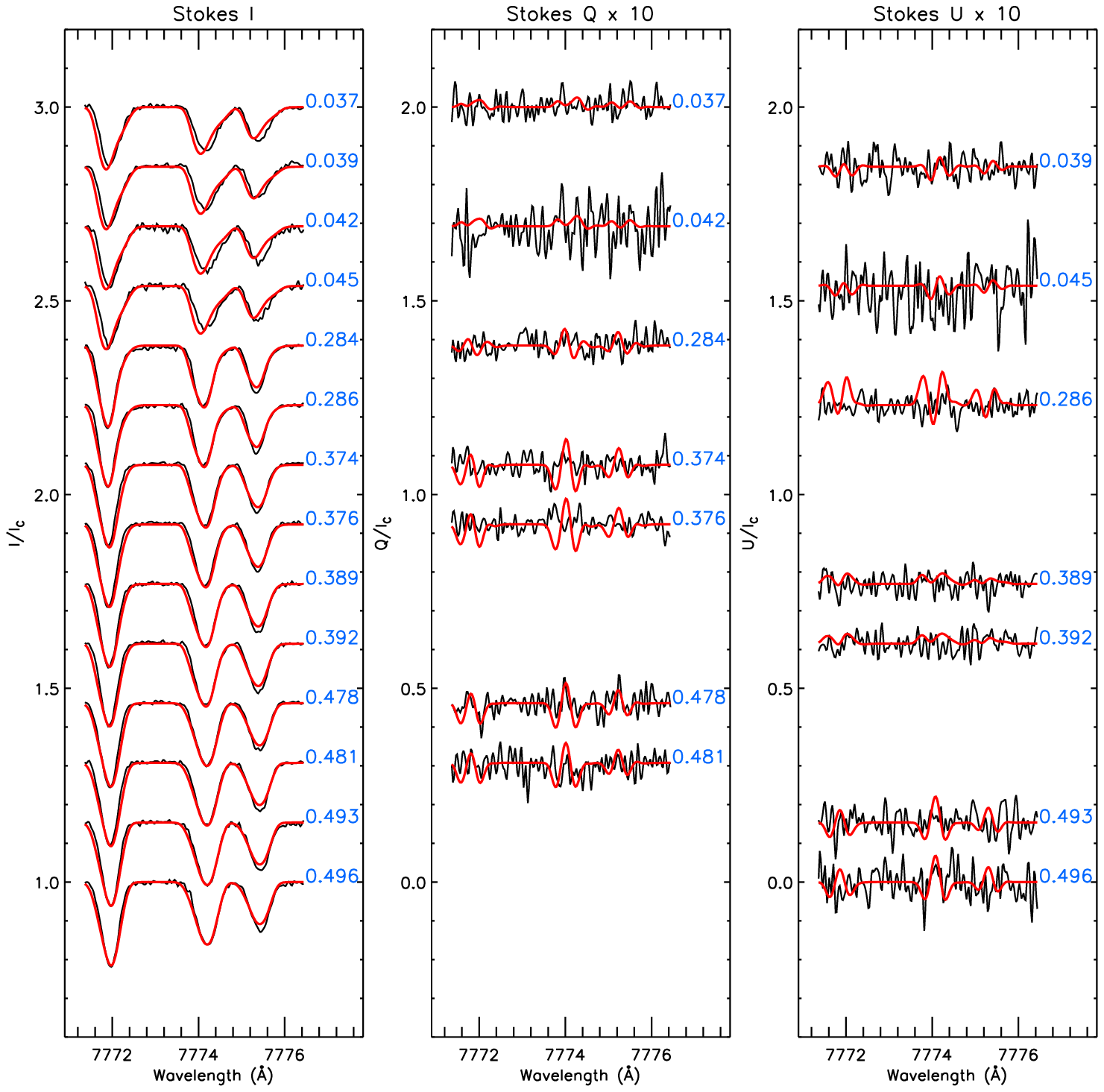}
 \caption{Comparison between observed (black lines) and synthetic Stokes $I$ and $Q/U$ parameters (red lines) of the oxygen triplet O~{\sc i} $\lambda\lambda$\,7772, 7774, 7775. Phases are indicated in blue.}
 \label{fig:mdi:qu}
\end{figure*}

As a final step in modeling the magnetic field of HD~184927, we have used our MDI model to compute linear polarisation profiles of individual spectral lines and compare them with the avaliable Stokes Q/U spectra. For this procedure we used the same lines as were used for computing the MDI maps -- He~{\sc i} $\lambda$\,6678 and the oxygen triplet O~{\sc i} $\lambda\lambda$\,7772, 7774, 7775. We detected no sign of linear polarisation in the observed profiles. The observations are reasonably consistent with the spectrum synthesis of the He line (except for phase 0.39 where the amplitude of the calculated profile is somewhat higher than the noise level). But for the O triplet, the predicted signatures have a higher amplitude than the observational noise for phases 0.28 and 0.37 (See Fig. \ref{fig:mdi:qu}). This may indicate a more complex field topology than inferred from the Stokes V analysis. Several authors previously reported a similar phenomenon occuring in some Ap stars \citep{bagn2001,kochukhov2004}.

The determination of \vsini\ described in Section \ref{sect:atmos} ignored the broadening introduced by the magnetic field. To test this assumption, we can use the magnetic field model derived from MDI to determine the magnetic broadening. We analysed the phase-averaged Si~{\sc iii} $\lambda$\,4567 and Fe~{\sc iii} $\lambda$\,5156 lines using the code Synmast.  We find a best-fit \vsini\ of 6.0-6.5 \kms, i.e. the impact of taking the magnetic field into account in the spectrum synthesis appears to be non-negligible, amounting to a reduction of \vsini\ by 3-4 \kms. However, we doubt that the \vsini\ obtained with this analysis is definitive. Due to the strong line profile variability, the average spectrum is not well reproduced with a simple spectrum synthesis assuming a homogeneous stellar surface. For example, fitting the phase-averaged synthetic profiles of the O triplet derived from the MDI modeling, we are unable to recover the \vsini=10 \kms with which profiles were computed. It appears that variability due to abundance spots introduces additional smoothing of mean line profiles.

It therefore seems likely that \vsini\ is in fact somewhat lower than that derived in Section \ref{sect:atmos}, although the exact value is rather uncertain. However, we note that a reduction of \vsini\ by several \kms\ does not influence the MDI results significantly because even with \vsini=10 \kms we are already in the regime where Doppler broadening does not dominate the line profiles. There is a marginal evidence that the Stokes I variations of the O triplet are best reproduced with \vsini$\sim$7 \kms. On the other hand, the He~{\sc i} $\lambda$\,6678 line seems to favour \vsini$\sim$9-10 \kms. In both cases, the change of the fit quality and resulting maps achieved by reducing \vsini\ is marginal.

\section{Magnetosphere}

In the presence of a sufficiently strong magnetic field, the interaction between the radiative winds of early-type stars and the magnetic field can lead to the formation of a magnetosphere \citep{bm1997, ud2002}. Wind plasma from opposite hemispheres is channeled along closed magnetic field lines to collide at the magnetic equator, producing X-rays at the wind shocks, along with relatively dense, cool clouds of stalled plasma. Magnetospheres can often be detected as emission in Balmer lines (e.g. \citealt{oks2012}), Paschen lines (e.g. \citealt{grun2012,eiken14}), and wind-sensitive ultraviolet resonance lines (e.g. \citealt{smithgroote2001, henrichs2005, oskinova2011}); X-ray over-luminosity (e.g. \citealt{oskinova2011, petit2013}); and photometric and broad-band linear polarization variations \citep{town2008, town2013, carc2013}. Due to co-rotation of the confined plasma with the magnetic field, these diagnostics are often (although not always) variable, with variations synchronized with the rotational period. 

The wind is considered to be magnetically confined if the ratio of magnetic to kinetic energy density in the wind, given by the magnetic wind confinement parameter $\eta_*$, is greater than unity \citep{ud2002}: 

\begin{equation}\label{etastar}
\eta_{*} \equiv \frac{B_{\rm eq}^{2}R_{*}^{2}}{\dot{M}v_{\infty}} 
\end{equation}

In calculating equation \ref{etastar} the equatorial magnetic field $B_{\rm eq}$ is used, as this is where the wind and the magnetic field most directly oppose one another; $R_*$ is the stellar radius; \mdot~is the mass-loss rate; and \vinf~is the wind terminal velocity. 

Following \cite{petit2013}, we use the theoretical recipe of \cite{vink2000} to calculate the wind parameters. With \mdot$=3.5\times 10^{-9}$, \msun{\rm yr}$^{-1}$ and  \vinf$= 894$ \kms, we find $\eta_*\sim 2.4^{+22}_{-1.1}\times 10^{4}$, where
the uncertainty is obtained by propagating uncertainties in $R_*$, $\teff$, and
$\log L$~through the calculations of \mdot, \vinf, and Eq. \ref{etastar}. Since $\eta_* > 1$, the wind is very likely magnetically confined.

The Alfv\'en radius \ra~(defined as the maximum extent of closed magnetic loops and thus defining the outer boundary of the magnetosphere; \citealt{ud2002}) can be determined heuristically from $\eta_*$ via

\begin{equation}\label{ralf}
\frac{R_{\rm A}}{R_*} \approx 0.3 + (\eta_* + 0.25)^{1/4}
\end{equation}

\noindent in this case yielding $R_{\rm A}\sim 12.7_{-1.4}^{+10} R_*$.

\cite{krtichka13} provide calculations of mass-loss rates and wind
terminal velocities specifically for chemically peculiar B-type stars.
However, if we adopt the tables of \cite{krtichka13} instead of
extrapolating the recipe of Vink et al., the wind-momentum is largely
unchanged: while \mdot~is smaller ($5.32 \time 10^{-10} M_\odot~{\rm
yr}^{-1}$), \vinf~is larger (4020 \kms). This then yields \ra $\sim 13 R_*$
and $\eta_* \sim 2.5 \times 10^4$: within the already existing
uncertainties.

An important consequence of magnetic wind confinement is rapid angular momentum loss
via the extended moment arm of the corotating plasma \citep{wd1967, ud2009}. The
angular momentum loss timescale $\tau_{\rm J}$ is given by \citep{petit2013}:

\begin{equation}\label{tauj}
\tau_{\rm J} = \frac{3}{2}f \tau_{\rm M}\left(\frac{R_*}{R_{\rm A}}\right)^2
\end{equation}

\noindent where $\tau_{\rm M} \equiv M_*/\dot{M}$ is the timescale over which
angular momentum is lost due to an unmagnetized wind, and $f$ is the moment of
inertia factor, which can be evaluated from the star's radius of gyration $\beta$ as
$f = \beta^2$. From the internal structure models of \cite{claret2004},
$f\sim0.0625$ for an 8 \msun~star of $\sim$7 Myr age, with little variation during
its previous evolution. Taking $\tau_{\rm M}$ to be constant, this then yields
$\tau_{\rm J} = 0.96$ Myr or 5.8 Myr, depending on whether the \citeauthor{vink2001}
or \citeauthor{krtichka13} mass-loss rates are used. 

We can use $\tau_{\rm J}$ and the rotation parameter $W$\,(the ratio of the rotation speed $V_{rot}$ to the orbital speed $V_{orb}$ at the equatorial surface radius $R_*$) to infer the star's spindown age $t_{\rm S}$
\citep{petit2013}:

\begin{equation}\label{tsmax}
t_{\rm S} = \tau_{\rm J}{\rm ln}\left(\frac{W_0}{W}\right)
\end{equation}

\noindent where $W_0$ is the initial rotation parameter. Assuming the star to have
been rotating at critical velocity, the maximum spindown age $t_{\rm s, max}$ can be
estimated: 3 Myr or 18 Myr, again depending on the adopted mass-loss rate. The
former estimate is easily accomodated within the 7 Myr age of the star.

\cite{petit2013} classified magnetospheres as either dynamical magnetospheres (DMs) or centrifugal magnetospheres (CMs). 

%

A CM is expected when \ra$>$\rk, where \rk~is the Kepler radius, the radius at which gravitational and centrifugal forces are in balance (\citealt{town2005a}; \citealt{ud2008}). \rk~is defined as

\begin{equation}\label{rkep}
R_{\rm K} \equiv \left(\frac{GM}{\omega^2}\right)^{1/3}
\end{equation}

\noindent where $G$ is the gravitational constant, $M$ is the stellar mass, and $\omega$ is the angular rotational velocity, which can be determined from $R_*$ and \vsini~if the inclination from the line of sight $i$ is known. Adopting the inclination, radius, and mass derived in Section 6, we find $R_{\rm K} = 7.5 \pm 3.3 R_*$.

Since \ra/\rk$> 1$, HD 184927 possesses a thin CM. However, unless \ra$>>$\rk~(typically, $\log_{10}{R_{\rm A}/R_{\rm K}} \ge 1$), emission is not typically observed in the optical part of the spectrum (\citealt{petit2013} Fig. 8; also Shultz et al., in prep.) In this case $\log_{10}{R_{\rm A}/R_{\rm K}} \sim 0.2$, so optical emission is not expected. 

Note that a star with a CM will still possess a DM, in the region $r<$\rk. So the star probably possess both a CM and DM.

X-ray and radio observations of HD 184927 are not yet available, so the magnetosphere can only be examined via optical and ultraviolet emission. 

\begin{figure*}
\centering
\begin{tabular}{cccc}
\includegraphics[width=4.3cm]{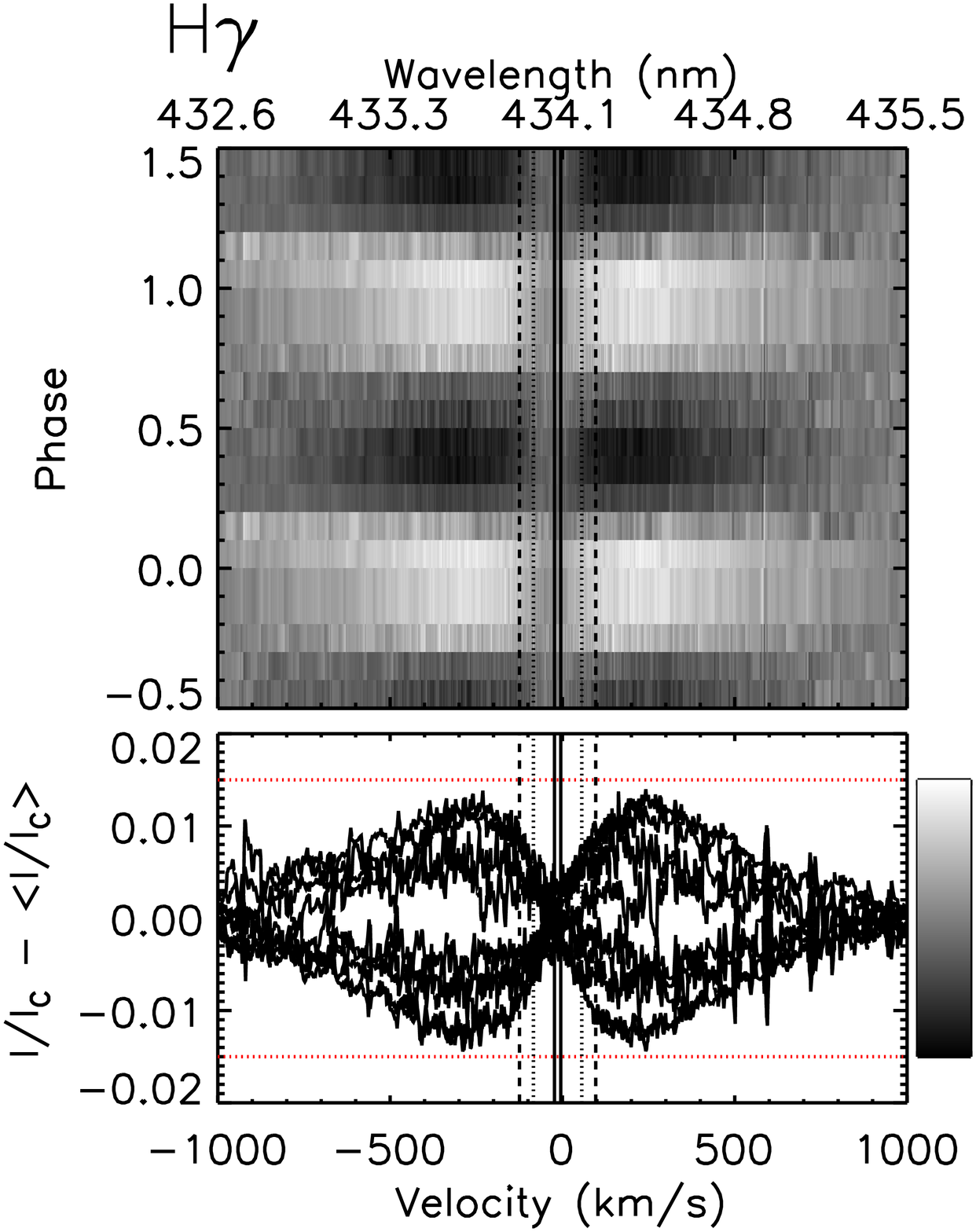} &
\includegraphics[width=4.3cm]{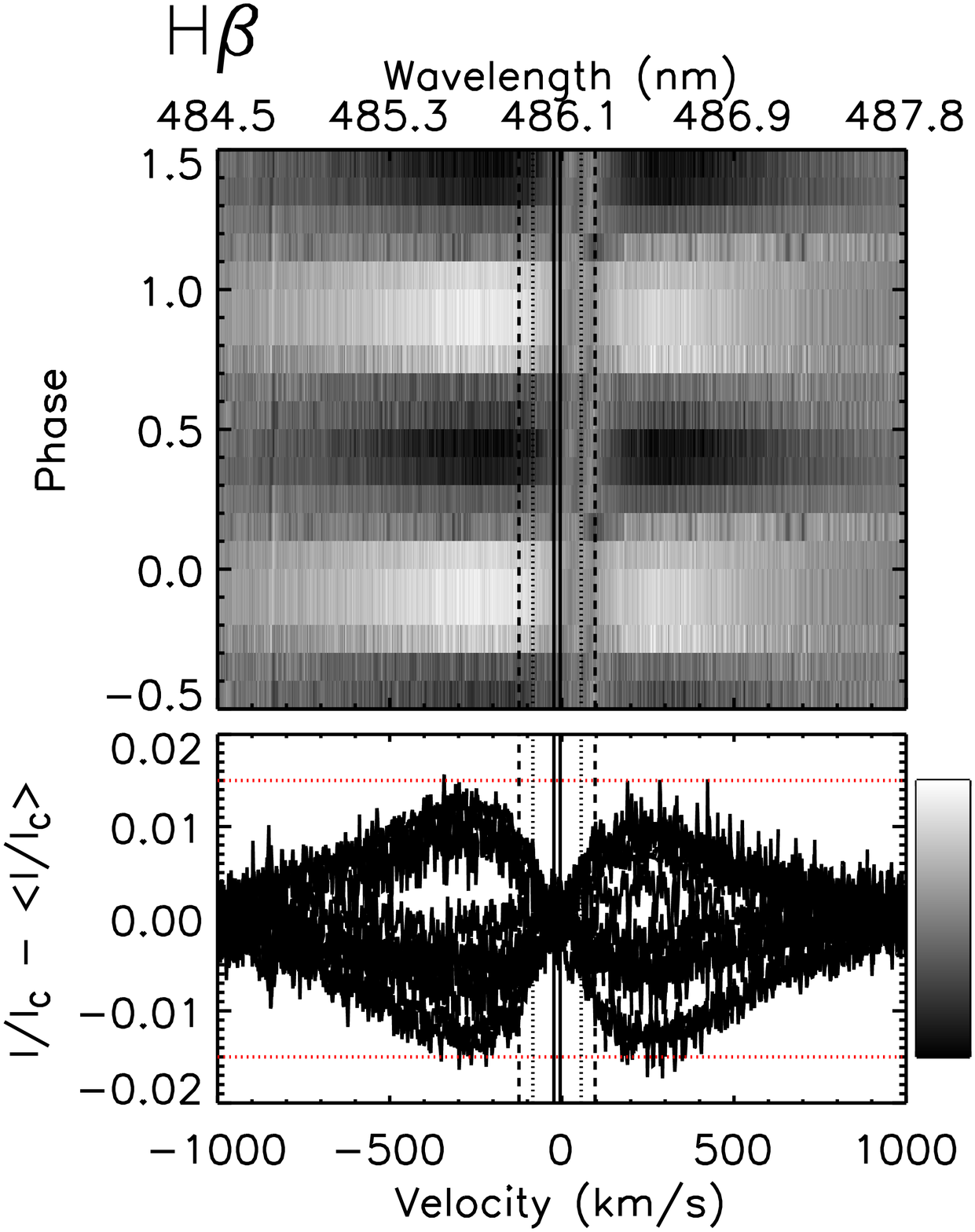} &
\includegraphics[width=4.3cm]{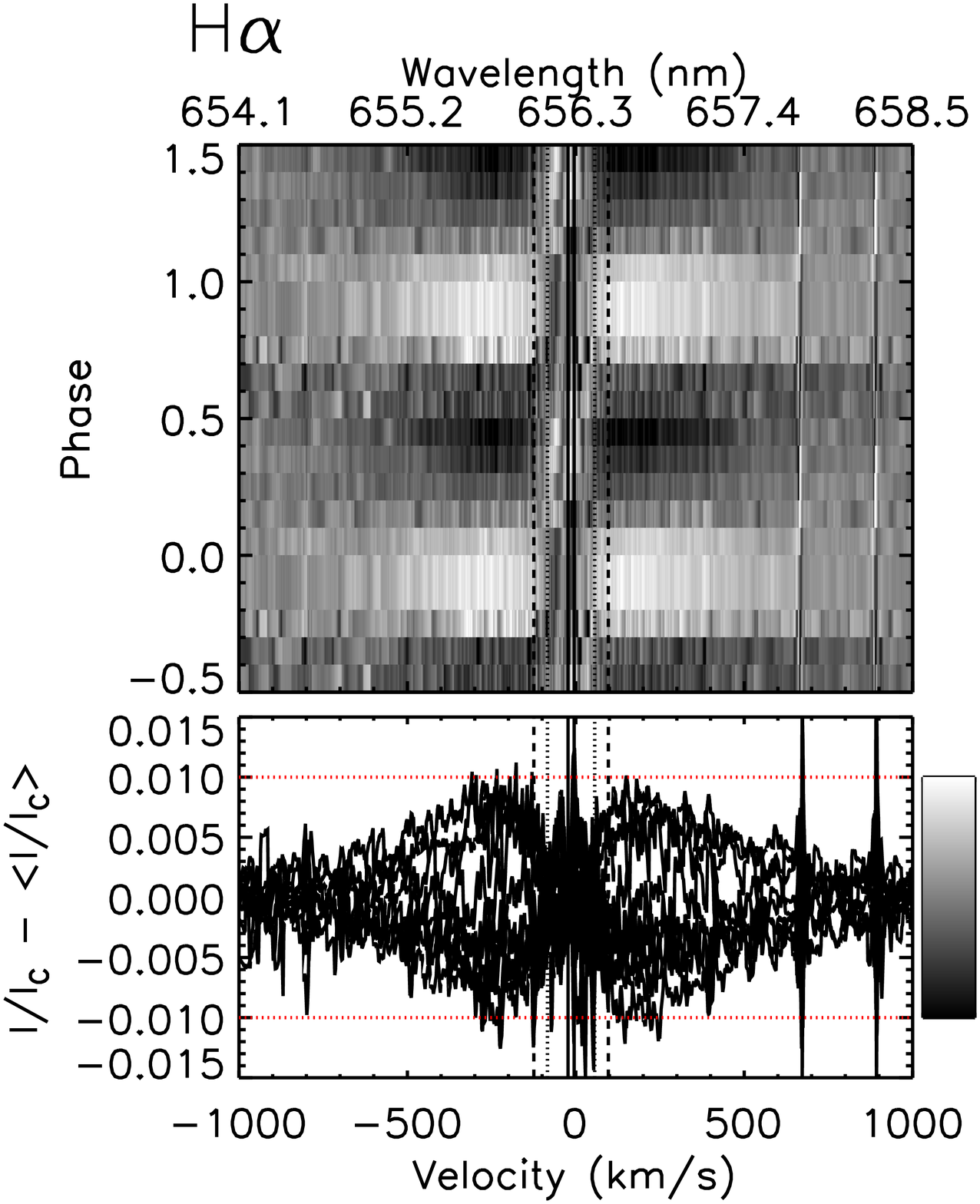} &
\includegraphics[width=4.3cm]{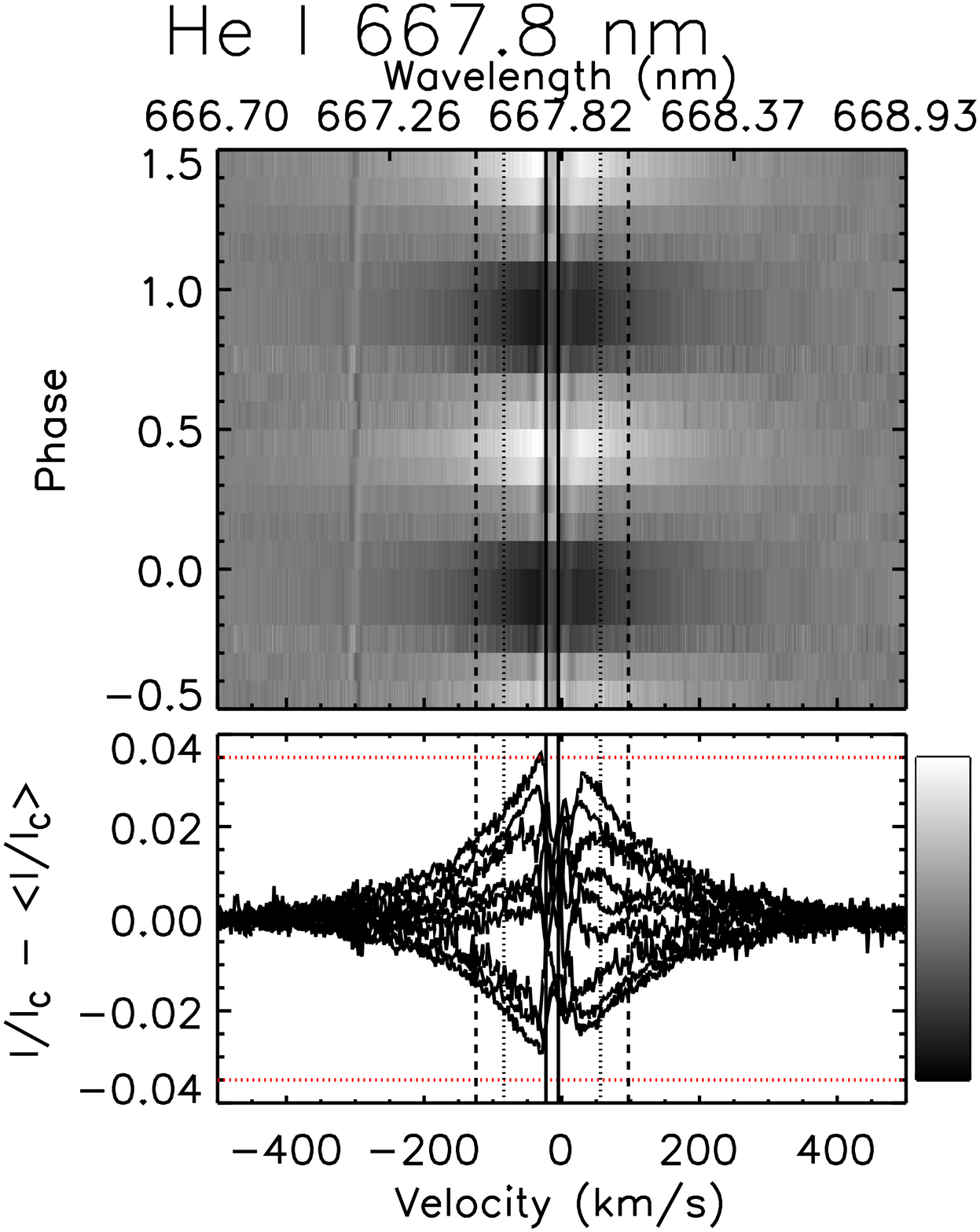} \\
\includegraphics[width=4.3cm]{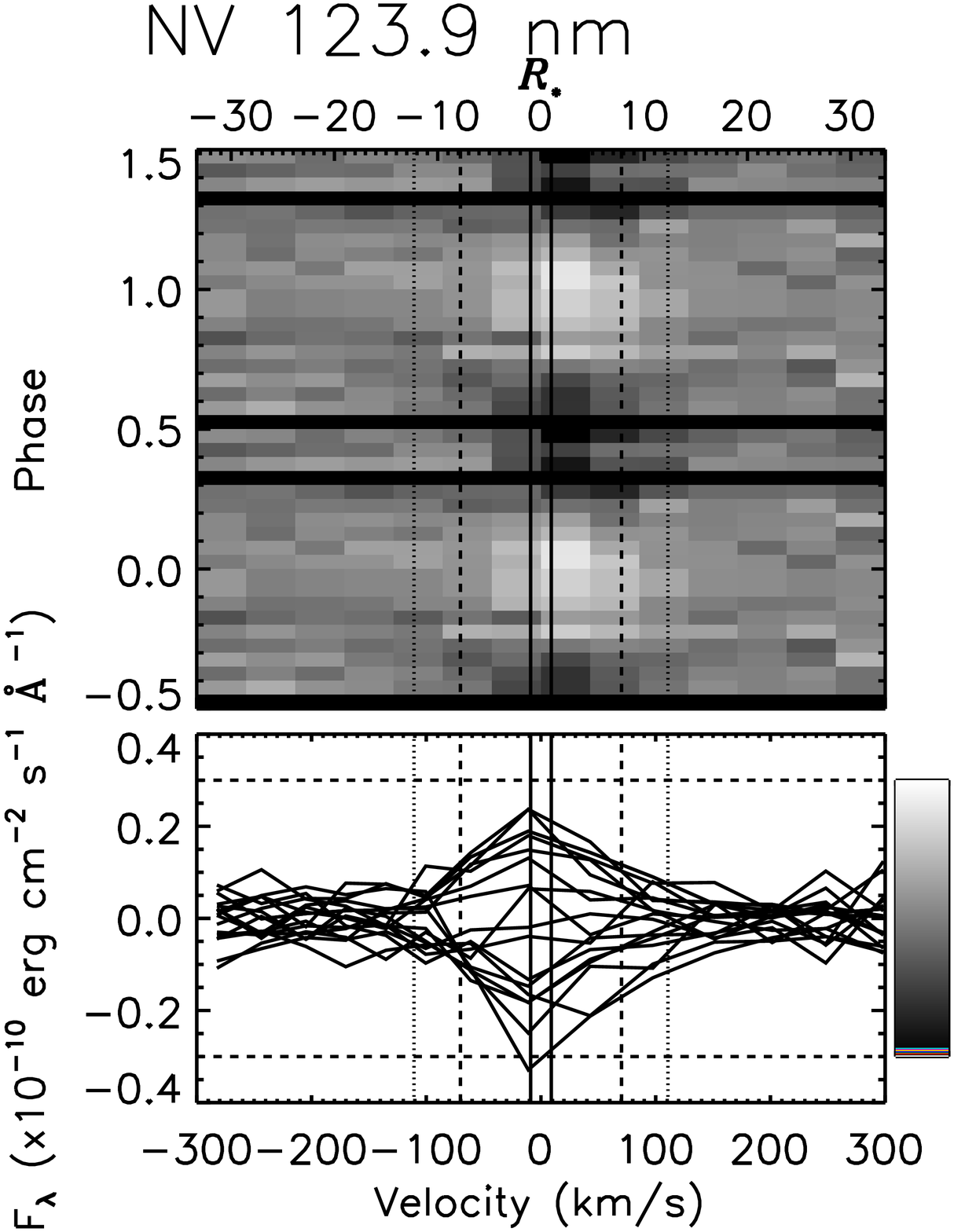} &
\includegraphics[width=4.3cm]{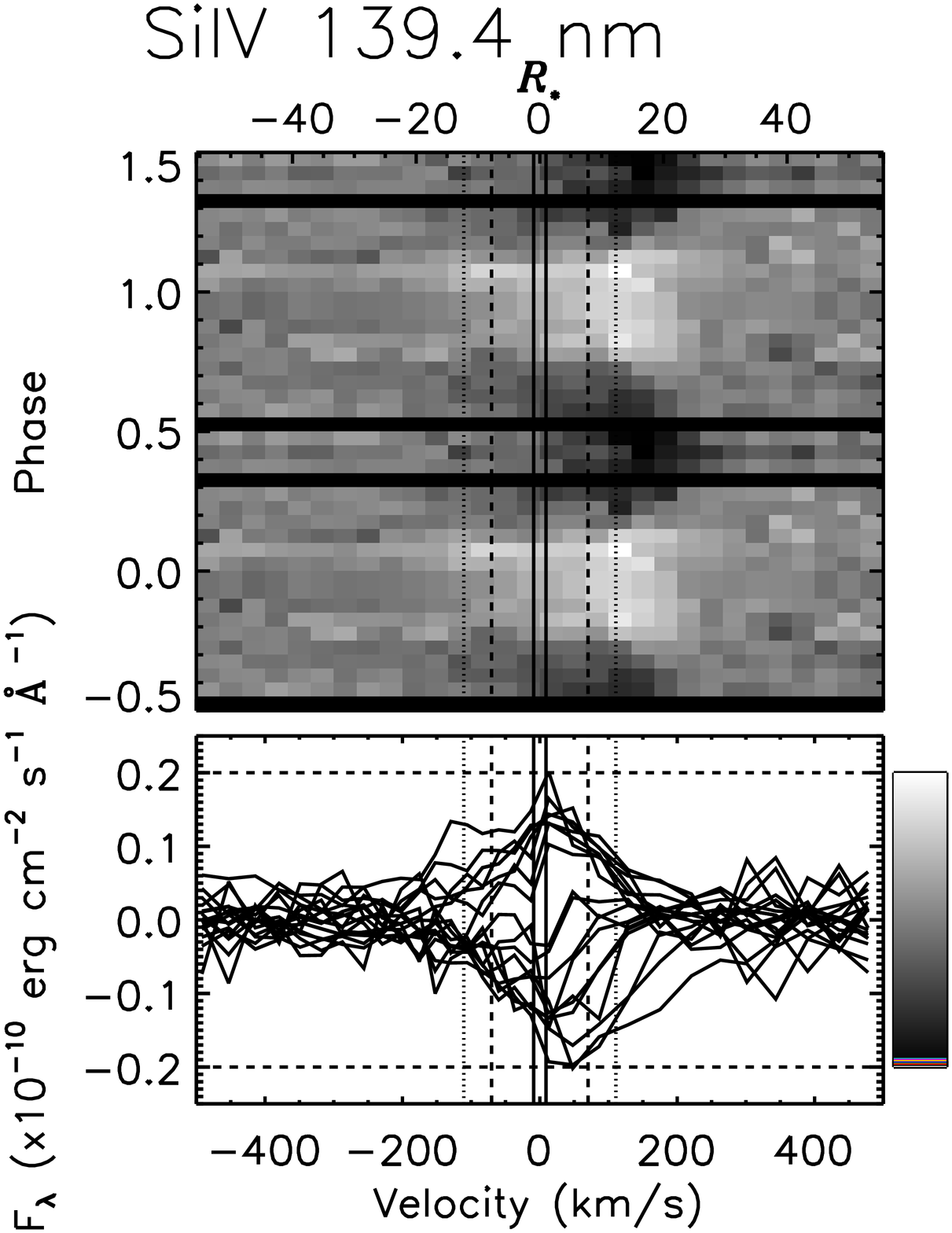} &
\includegraphics[width=4.3cm]{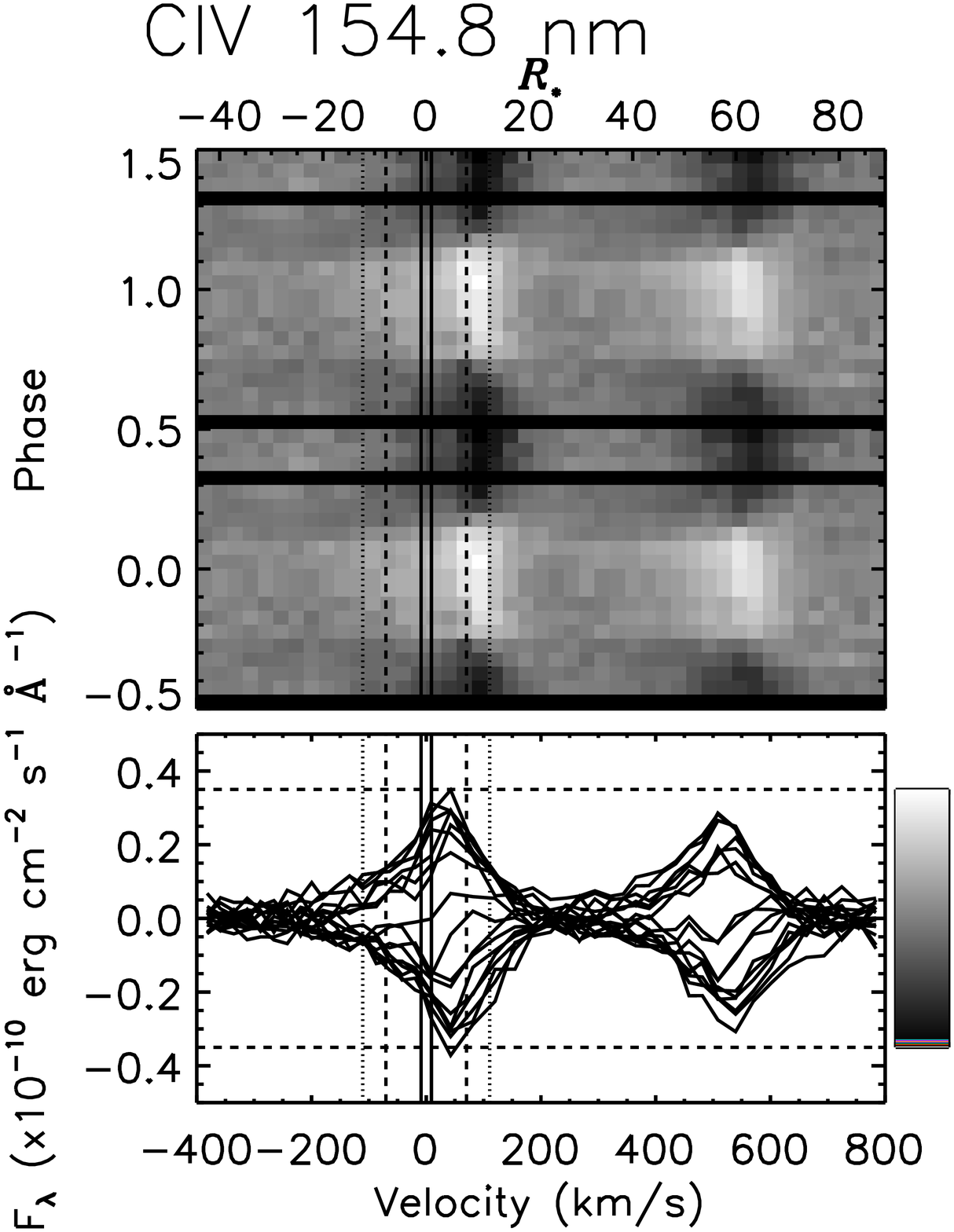} &  
\includegraphics[width=4.3cm]{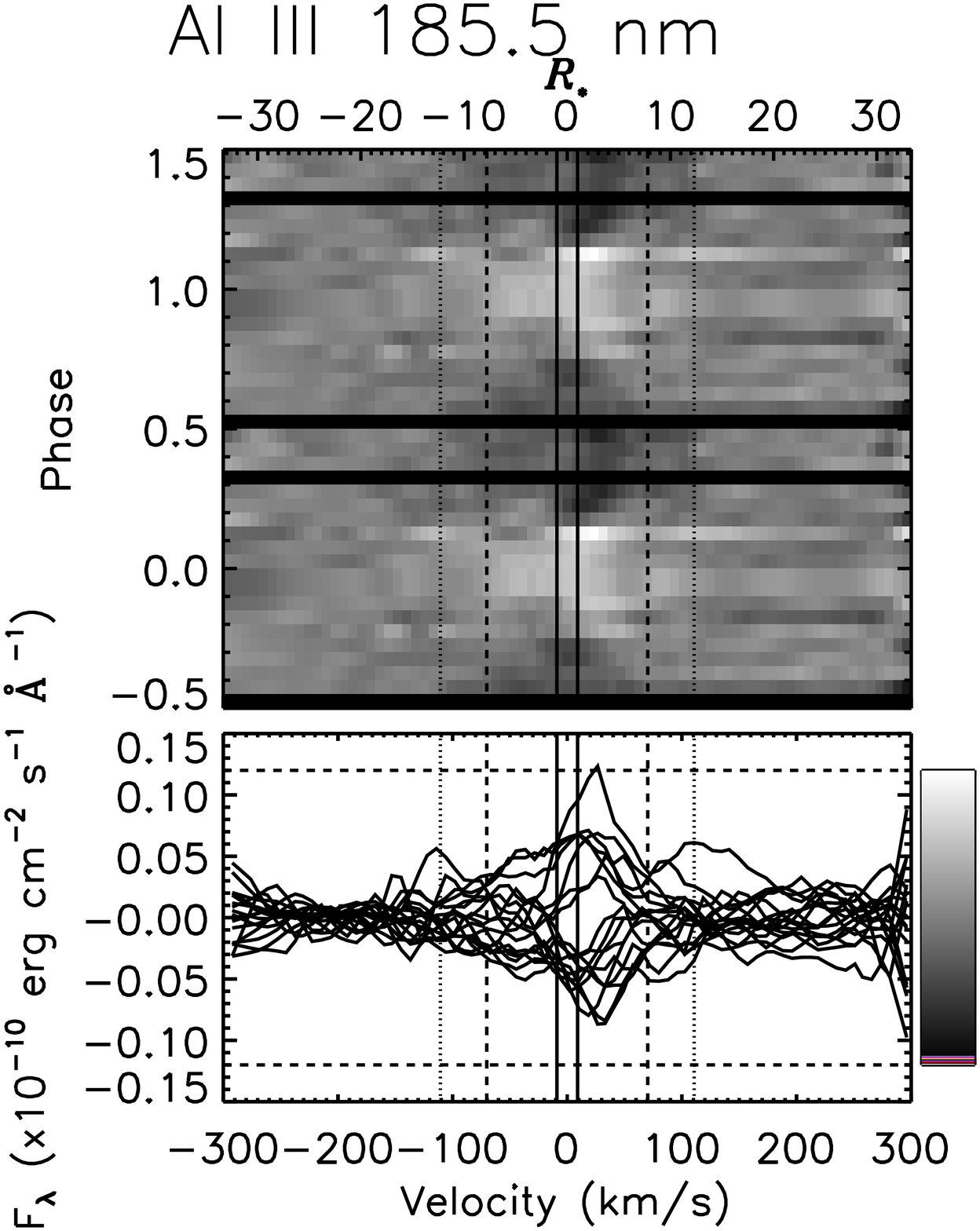} \\
\end{tabular}
\caption{Dynamic spectra of optical Balmer and He~{\sc i}~$\lambda$\,6678 and UV spectral lines of HD~184927. Solid vertical lines indicate $\pm$\vsini; dotted lines indicate \rk; dashed lines indicate \ra;red horizontal lines indicate
the mapping of the colour bar to the residual flux. Bottom panels show residual intensity after comparison to a mean spectrum. {\em Above}:  (left--right) H$\gamma$, H$\beta$, H$\alpha$, and He {\sc i}~$\lambda$\,6678. {\em Below}:  Selected ultraviolet resonance lines. As variability in these lines is almost certainly circumstellar in origin, the top abscissa of the UV dynamic spectra is given in units of stellar radii rather than wavelength. There is almost no difference between the Balmer lines, and a strong anti-correlation exists between Balmer and He variability. Balmer line variability is also uncorrelated with either \ra~or \rk, while variability within the He and UV lines is approximately bounded by $\pm$\ra.}
\label{dynspecs}
\end{figure*}

The Balmer lines are variable at about 2\% of the continuum (see first two panels of Fig. \ref{dynspecs}). The magnitude of variability is similar for all Balmer photospheric lines (compare H$\gamma$ to H$\alpha$), whereas for circumstellar emission the variability in H$\alpha$ should be much more pronounced than in other H lines. The variability is not bounded by \ra, as expected for a magnetosphere. Finally, Balmer line variability is almost a mirror image of the variability in He lines (e.g. He {\sc i} 667.8 nm, right panel of Fig. \ref{dynspecs}). All of this suggests that this variability is photospheric in origin, possibly a consequence of displacement of H by He. We conclude that there is no signature of HD 184927's magnetosphere detected in the Balmer lines of the ESPaDOnS spectra. The variability in the Paschen series resembles that in the Balmer series. Just as with the Balmer series, there's every reason to suspect this to be photospheric in origin. 

\begin{figure}
\centering
\includegraphics[scale=0.29]{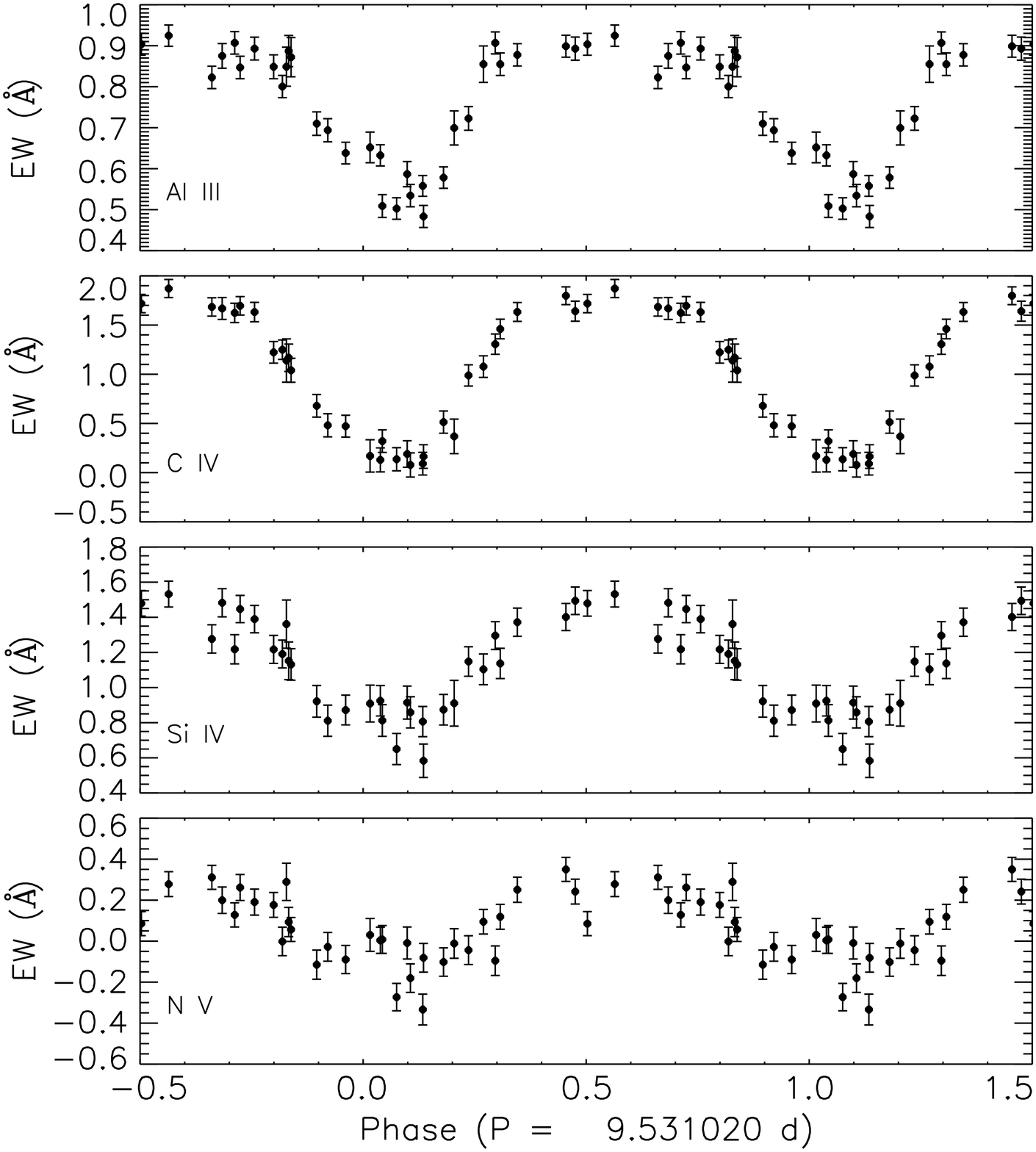} 
\caption{EW variations of wind-sensitive ultraviolet doublets.}
\label{uv_ew}
\end{figure}

Ultraviolet spectra are more useful in this regard. The variability of the IUE UV spectra of HD 184927 is well known (e.g. \citealt{barker1982}, \citealt{brown1987}). Variability is present in numerous lines; examples are shown for Al {\sc iii}, C {\sc iv}, Si {\sc iv}, and N {\sc v} in Figs. \ref{dynspecs} and \ref{uv_ew}. This is especially significant in the case of N {\sc v}, as the high ionization potential (92 eV) of this doublet requires a higher temperature than is available in the photosphere. The presence of this line is a direct consequence of `superionization' due to X-ray production in the magnetically confined wind shock (e.g. \citealt{oskinova2011}). 

Essentially the same variability pattern is present in all four doublets, as is demonstrated in both the dynamic spectra (Fig. \ref{dynspecs}, bottom panels) and the composite EW measurements (i.e., combined measurements for both lines in the doublet; Fig. \ref{uv_ew}). There are telling differences between the UV lines and the H and He lines. The variability of the latter is entirely symmetrical about the $v=0$ \kms\, while the former (especially C {\sc iv} and Al {\sc iii}) are not symmetrical. UV line variability is essentially bounded by \ra~(see Fig. \ref{dynspecs}), as
expected for emission formed within a magnetosphere corotating with the
stellar surface.


\cite{smithgroote2001} modeled absorption and emission in a variety of low- and high-excitation UV lines, determining the temperature, turbulence, and column density for each line. Comparing results for lines with the lowest and highest excitation potentials, they found temperatures varying from $\sim$ 15 kK to $\sim$45 kK, turbulent velocities from 20--50 \kms, and column densities from $10^{23}$ to $10^{22} {\rm cm}^{-2}$: that is, the higher-energy lines were best fit using lower column densities, greater turbulence, and higher temperatures. Turbulence may account for the presence of variability at velocities greater than those accounted for by \ra~(see Fig. \ref{dynspecs}). 

The presence of significant turbulence, together with the strong red-shifted emission peaks at low velocities (interpreted by \citeauthor{smithgroote2001} as downflows), suggest that the UV resonance lines examined here are formed relatively close to the star within the DM, rather than within its very thin CM. Assuming the EW increases can be fully explained by
occultations of the star by the circumstellar plasma torus, their duration can be used to infer the distance of the cloud from the photosphere under the assumption of strict corotation. The occultations are quite long (lasting approximately 0.5 of a rotational cycle, see Fig. \ref{uv_ew}), suggesting that the clouds are located very close to the photosphere. As this is well within \rk, this supports the argument that the UV variations are produced primarily within the DM.

\section{Discussion \& conclusions}

In this paper we have examined an extensive collection of spectropolarimetric and spectroscopic observations of the He strong star HD 184927. We used previously published magnetic field measurements together with our own measurements of the longitudinal magnetic field and EW to improve the rotation period of the star to $P =9.53102\pm0.0007$~d - a value ten times more precise than that previously published by \cite{wade97}. 

We also examined the spectral line EW and profile variations. We find that lines of many elements are variable. However, whereas most elements show maximum EW near rotational phase 0.5, lines of He vary in the opposite sense, with maximum strength near phase 0.0. The line variations, completed by NLTE spectrum synthesis assuming a patchy surface distribution of He, supports the view that many elements are distributed non-uniformly across the surface of the star.

We performed an extensive analysis of the spectrum and fluxes of HD 184927, modelled using the TLUSTY NLTE model atmosphere code. This allowed us to improve the precision and accuracy of the physical characteristics of the star. Some of these properties - most notably the luminosity - are substantially revised relative to those presented in earlier investigations of this star. 

Motivated by important differences between the observed characteristics of the longitudinal field curves derived from H lines and from LSD profiles, we performed a detailed examination of the dependence of the longitudinal field curve on chemical species. We find large differences between the variations of different elements. While the field curve measured from H lines varies from $\sim 0-2$~kG, that of He varies from only 0-1~kG, and those of N, O, Si and Fe vary from about $\sim -700$~G to 1~kG. 

These large differences in the longitudinal field variations - which we interpret as a consequence of the different distributions of these elements on the stellar surface - could potentially have important implications for the determination of the magnetic field geometry. In our analysis, we derive a rotation axis inclination $i=25\degr$ using the stellar radius, rotational period and $v\sin i$. It is not uncommon to assume a dipole field and to compute the best combination of the magnetic obliquity $\beta$ and dipole polar strength $B_{\rm d}$ by fitting the field curve \citep[e.g.][]{wade97}.

 \begin{figure}
 \includegraphics[scale=0.6]{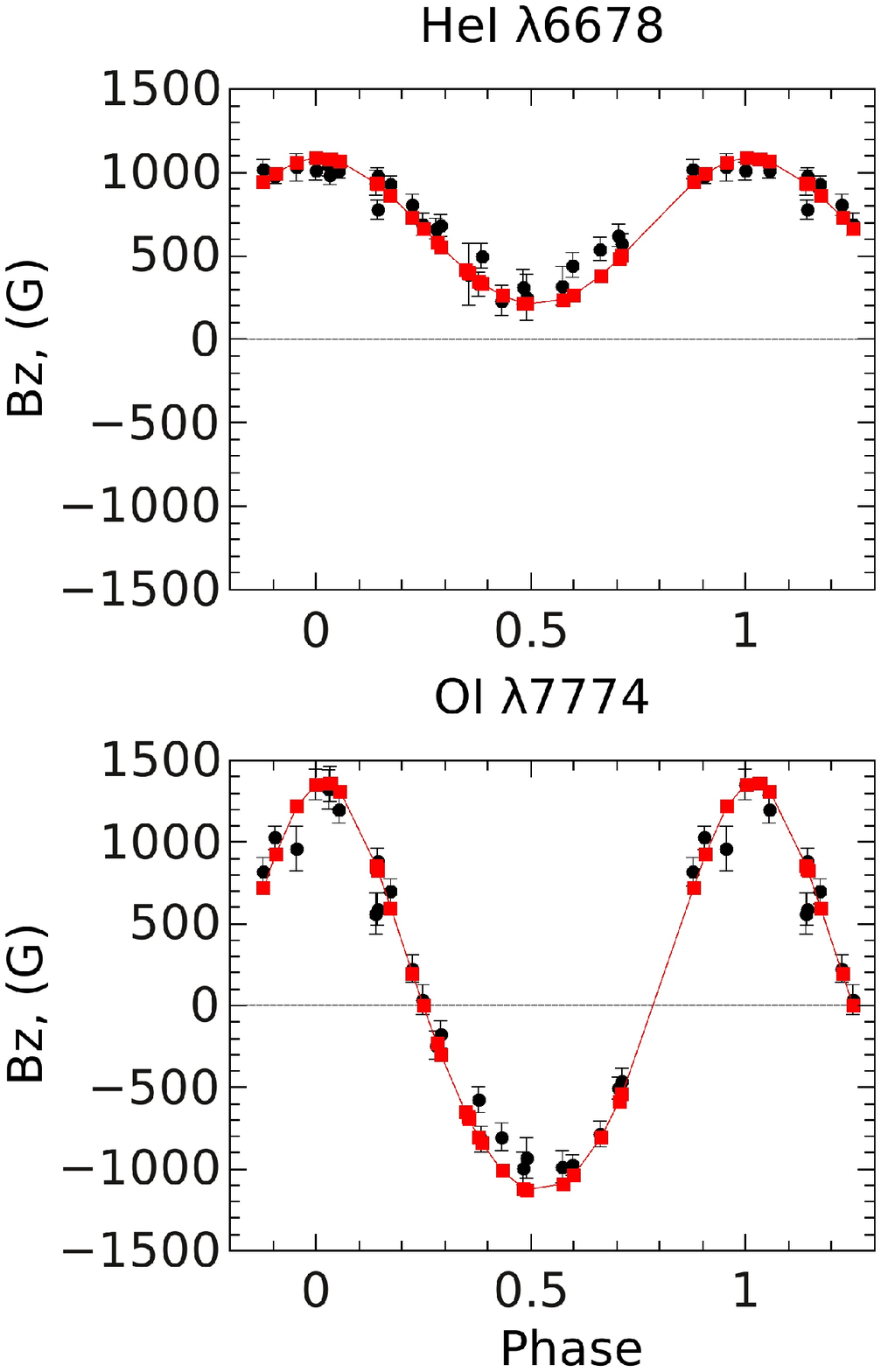}
  \caption{Comparison between observed longitudinal magnetic curves (black dots) and variations computed from the magnetic and abundance maps derived from MDI (red curve).  \label{fig:mdi:compare}
}
 \end{figure}

For example, if we fit the H longitudinal field curve, we obtain $\beta\simeq 65\degr$ and $B_{\rm d}\simeq 8.5$~kG. On the other hand, fitting the field curve measured from the He lines gives the same obliquity but a polar field strength about half as large. Finally, fitting the field curve from O lines yields an intermediate polar field ($B_{\rm d}\simeq 7$~kG) and a substantially greater obliquity ($\beta\simeq 87\degr$). 

\cite{wade97} proposed two difference models of the magnetic field geometry depending on the adopted rotational velocity ($B_{\rm d}\simeq 9.7$~kG, $B_{\rm oct}\simeq 2.0$~kG, $i=29\degr$, $\beta=76\degr$ for $v\sin i = 17$~\kms and $B_{\rm d}\simeq 13.7$~kG, $B_{\rm q}\simeq -9.0$~kG, $B_{\rm oct}\simeq 12.0$~kG, $i=20\degr$, $\beta=81\degr$ for $v\sin i = 12$~\kms). They added quadrupolar and octupolar magnetic moments in their calculations, but noted that this modification does not change the computed longitudinal field significantly. However, they were unable to distinguish between the two models based on the data they had and concluded that the magnetic field configuration of HD 184927 lies somewhere between these two extremes.

In our investigation, we used Magnetic Doppler Imaging to simultaneously model the magnetic field and chemical abundance distributions of the elements He and O. In Fig.~\ref{fig:mdi:compare} we once again illustrate the longitudinal field variations measured from the He~{\sc i} $\lambda 6678$ and O~{\sc i} $\lambda 7774$ lines. Overplotted are the calculated variations obtained from the synthetic Stokes $I$ and $V$ profiles characterizing the MDI model fit to the data. Given that the magnetic field maps derived from these two lines are nearly identical, the excellent agreement illustrated in Fig.~\ref{fig:mdi:compare} implies that the large differences in their longitudinal field curves can be fully explained as the consequence of a single, unique magnetic field distribution in combination with element-specific chemical abundance distributions.

Our investigation of HD 184927 represents the first complete, modern analysis of one of the most slowly rotating He strong stars. Analyses such as this represent important contributions to a systematic understanding of the physics of magnetic massive stars.

\section{Acknowledgements}
This investigation is based on MiMeS Large Program observations obtained at the Canada-France-Hawaii Telescope (CFHT) which is operated by the National Research Council of Canada, the Institut National des Sciences de l'Univers (INSU) of the Centre National de la Recherche Scientifique of France, and the University of Hawaii. Some of the data presented in this paper were obtained from the Multimission Archive at the Space Telescope Science Institute (MAST). STScI is operated by the Association of Universities for Research in Astronomy, Inc., under NASA contract NAS5-26555. Support for MAST for non-HST data is provided by the NASA Office of Space Science via grant NAG5-7584 and by other grants and contracts.

IY and TS thank the Russian Foundation for Basic Research for financial support of this study (projects 12-02-31246-mol-a, 14-02-31780). GAW is supported by a Discovery Grant from the Natural Sciences and Engineering Research Council (NSERC) Canada.

OK is a Royal Swedish Academy of Sciences Research Fellow, supported by the grants
from the Knut and Alice Wallenberg Foundation, Swedish Research Council, and the
G\"oran Gustafsson Foundation. The computations presented in this paper were
performed on resources provided by SNIC through Uppsala Multidisciplinary Center for
Advanced Computational Science (UPPMAX).

Authors thank Dr. Ivan Hubeny for provision of the \verb|SYNPLOT| and \verb|SYN_ABUND| codes.
\bibliography{hd184927}{}

\label{lastpage}

\end{document}